\documentclass{imsart}

\RequirePackage{amsthm,amsmath}
\RequirePackage[numbers]{natbib}
\RequirePackage[colorlinks,citecolor=blue,urlcolor=blue]{hyperref}
\RequirePackage{graphicx}
\RequirePackage{amsmath,amssymb}
\RequirePackage{array}
\RequirePackage{threeparttable}
\RequirePackage{multirow}
\RequirePackage{subfigure}
\RequirePackage{color}

\startlocaldefs
\theoremstyle{plain}
\newtheorem{theorem2}{Theorem}[section]
\newtheorem{lemma1}{Lemma}[section]
\newtheorem{remark1}{Remark}[section]
\newtheorem{assumption}{Assumption}

\endlocaldefs

\begin{document}

\begin{frontmatter}

\title{Efficient Change Point Detection and Estimation in High-Dimensional Correlation Matrices\support{Zhaoyuan Li’s research is partially supported by a National Natural Science Foundation of China grant (No. 11901492).}}
\runtitle{Efficient Change Point Detection and Estimation in HD Correlation Matrices}

\begin{aug}
\author{\fnms{Zhaoyuan} \snm{Li}\textbf{}\corref{}\ead[label=e1]{lizhaoyuan@cuhk.edu.cn}}
\and
\author{\fnms{Jie} \snm{Gao}\ead[label=e2]{jiegao2@link.cuhk.edu.cn}}

\address{School of Data Science \\  The Chinese University of Hong Kong, Shenzhen \\ Shenzhen, China, 518172\\ \printead{e1,e2}}

\runauthor{Li et al.}
\end{aug}

\begin{abstract}
	This paper considers the problems of detecting a change point and estimating the location in the correlation matrices of a sequence of high-dimensional vectors, where the dimension is large enough to be comparable to the sample size or even much larger. A new break test is proposed based on signflip parallel analysis to detect the existence of change points. Furthermore, a two-step approach combining a  signflip permutation dimension reduction step and a CUSUM statistic is proposed to estimate the change point's location and recover the support of changes. The consistency of the estimator is constructed. Simulation examples and real data applications illustrate the superior empirical performance of the proposed methods. Especially, the proposed methods outperform existing ones for non-Gaussian data and the change point in the extreme tail of a sequence and become more accurate as the dimension $p$ increases. Supplementary materials for this article are available online. 
	
\end{abstract}

\begin{keyword}[class=MSC]
\kwd[Primary ]{62M10}
\kwd{62H15}
\kwd[; secondary ]{62P20}
\end{keyword}

\begin{keyword}
	\kwd{Change point detection}
	\kwd{Change point estimation}
	\kwd{Signflip parallel analysis}
	\kwd{Correlation matrix}
	\kwd{CUSUM}
\end{keyword}


\tableofcontents

\end{frontmatter}

\section{Introduction}
	Change point detection is a classical statistical problem aiming to detect if there is a change in the mean, covariance structure, and distribution along a sequence of time-ordered observations. It has been an active field of research in many scientific fields, such as quality control \citep{page1954continuous}, financial market analysis \citep{wied2017nonparametric}, genetics and medicine \citep{chen2012parametric}, psychopathology \citep{rosenfield2010change}, signal processing \citep{jiruska2013synchronization} and machine learning \citep{ cao2018sequential}. 
	
	Driven by a wide range of modern scientific applications, change point inference in high-dimensional data is of significant current interest. A vast part of the existing literature focuses on detecting changes in the mean \citep{jirak2015uniform,cho2015multiple,wang2018high} and in the covariance structure \citep{aue2009break,dette2016detecting,avanesov2018change,dette2020estimating,covariance2}. The literature on detecting structural breaks in the correlation matrix is relatively scarce. The correlation coefficient is a standard measure for linear dependency among variables, and correlation analysis is reliable only when the data has no change in correlation during the observation period. Moreover, in a multivariate setting, aside from changes in univariate features, important events are often marked by abrupt correlation changes \citep{cabrieto2018capturing}. 
	
	
	Change point analysis of correlation has been well studied in the conventional low-dimensional setting, i.e., the dimension is fixed while the sample size becomes large. 
	\cite{wied2017nonparametric} constructed a bootstrap variance matrix estimator to detect a change in the correlation matrix.  \cite{cabrieto2018capturing} proposed a Gaussian kernel-based change point detection method (KCP). Compared to CUSUM-type methods, KCP can locate multiple change points simultaneously.
	Change point analysis in time series has been studied by \cite{killick2013wavelet} and \cite{dette2019change} and the references therein.
	However, in the high-dimensional setting, where the dimension can be much larger than the sample size, the methods constructed under the low-dimensional setting either perform poorly or are not even well defined. For example, the method proposed by \cite{wied2017nonparametric} is unstable when the dimension is not small relative to the sample size due to the (near) singularity of the variance estimator. To avoid variance estimation,
	\cite{choi2020self} proposed a break test based on the self-normalization method, which applies to high-dimensional data. However, there is no estimation method proposed to locate the change point. \cite{dette2020estimating} proposed a two-stage approach based on bootstrap to estimate a change point in the high-dimensional covariance structure. This method can be applied to high-dimensional correlation matrices, as correlation can be treated as a particular case of the covariance matrix. However, this method is designed to estimate the location when the change point exists in the middle of a sequence of data, which is restrictive in real applications.  
	
	This paper proposes a new test based on signflip parallel analysis to detect the existence of changes in the correlation structure. Moreover, we construct a two-step approach combining a signflip permutation dimension reduction step and a CUSUM statistic to estimate the location of change point and simultaneously identify corresponding changing components in the correlation matrix. The theoretical properties analysis is conducted, and the consistency of the proposed estimator is established. In addition, by combing the Synthetic Minority Oversampling Technique (SMOTE), the estimation accuracy is enhanced when the change point exists in the extreme tail of a data sequence. We examine the numerical performance of the proposed detection and estimation methods using both simulated and real datasets. The numerical results show that the proposed methods significantly outperform the existing ones for non-Gaussian data and the change point existing in the extreme tail of a sequence. More importantly, as the dimension $p$ increases, the proposed methods become more accurate while existing methods maybe not. 
	
	The remaining sections are organized as follows. In Section 2, we present the first main result of the paper. A new break test is proposed based on  signflip parallel analysis to detect change points in the correlation structure of high-dimensional data. In Section 3, we present the second main result of the paper. After a dimension reduction step based on signflip permutation, a CUSUM statistic is used to estimate the location of the change point. At the same time, the components of the correlation matrix leading to the change point are identified. The consistency of this estimator is established. In Section 4, an algorithm combing the proposed estimation procedure and SMOTE is designed to locate the change point in the extreme tail of a data sequence. We evaluate the finite performance of two proposed methods compared to several existing methods by a detailed simulation study in Section 5. Real data analyses are carried out in Section 6. Some discussions are offered in the last conclusion section. Due to the limited space, we relegated the proofs and some numerical results to supplementary materials. 
	
	\section{Detection of Change Point}
	
	Given a sequence of data $\boldsymbol{y}_1, \ldots, \boldsymbol{y}_{t_0}, \boldsymbol{y}_{t_0+1}, \ldots, \boldsymbol{y}_T\in \mathbb{R}^p$, as a first step, one may be interested in testing the existence of change points in the correlation structure. In this section, we aim to test for consistency of the correlation matrices of observations $\{\boldsymbol{y}_t \}_{1\leq t\leq T}$:
	\begin{equation}\label{hypothesis_detection}
		H_0: corr(\boldsymbol{y}_1)=\cdots=corr(\boldsymbol{y}_T), \quad \text{v.s.} \quad H_1: \text{not} \ H_0.
	\end{equation}
	The null hypothesis $H_0$ indicates the consistency of the correlation matrices during the observation period. The alternative hypothesis allows for one or more change points in the correlation structure. When there exists a change point, the parameter $t_0$ defines the true change point, that is, the first $t_0$ samples $\boldsymbol{y}_1, \ldots, \boldsymbol{y}_{t_0}$ have a common correlation matrix $R_1=(\rho_1 (i,j)), i,j=1,\ldots, p$, while the last $(T-t_0)$ samples $\boldsymbol{y}_{t_0+1}, \ldots, \boldsymbol{y}_T$ have another common correlation matrix $R_2=(\rho_2 (i,j)), i,j=1,\ldots, p$, and $R_1 \neq R_2$.
	
	For $t=1, \ldots, T$, denote by $\boldsymbol{y}_t=(y_{t1}, \ldots, y_{tp})'$ the $t$-th observation, let $\bar{\boldsymbol{y}}=\frac{1}{T}
	\sum_{i=1}^{T}\boldsymbol{y}_i =(\bar{y}_1, \ldots, \bar{y}_p)'$ be the sample mean, $S_n=\frac{1}{T-1}\sum_{i=1}^T(\boldsymbol{y}_i-\bar{\boldsymbol{y}})(\boldsymbol{y}_i-\bar{\boldsymbol{y}})'$ the sample covariance matrix with diagonal elements diag(D). Define $\boldsymbol{x}_t={D}^{-1/2}(\boldsymbol{y}_t-\bar{\boldsymbol{y}})$ as the vector of standardized observations. Given a time point $t$, define the sample correlation matrices
	\[ \left( \hat{\rho}_1^t(i,j) \right)_{p\times p}:= \hat{R}_1^t=\frac{1}{t}\sum_{k=1}^t \boldsymbol{x}_k \boldsymbol{x}'_k, \]
	\[ \left(\hat{\rho}_{t+1}^T(i,j)  \right)_{p\times p}:=\hat{R}_{t+1}^T=\frac{1}{T-t}\sum_{k=t+1}^T\boldsymbol{x}_k \boldsymbol{x}'_k. \]
	In order to quantify the changes in correlations after and before $t$, a natural approach is to compare the sample correlations $\{ \hat{\rho}_1^t(i,j)\}$ and $\{\hat{\rho}_{t+1}^T(i,j) \}$, and we consider the following $p(p-1)/2$-dimensional vector
	\begin{eqnarray}\label{Vt}
		\left(\left[\hat{\rho}_1^t(i,j)-\hat{\rho}_{t+1}^T(i,j)\right]^2 \right)_{1\le i< j \leq p}:= \boldsymbol{v}_t 
	\end{eqnarray}
	of the squared (componentwise) differences of the elements of the sample correlations. For large $p$, it can be calculated efficiently by
	\begin{equation*}
		\begin{aligned}
			\boldsymbol{v}_t
			&= \left( \frac{1}{t}\sum\limits_{k=1}^t vecho(\boldsymbol{x}_k \boldsymbol{x}'_k) - \frac{1}{T-t}\sum\limits_{k=t+1}^T vecho(\boldsymbol{x}_k \boldsymbol{x}'_k) \right) \\
			&\quad \circ \left(\frac{1}{t}\sum\limits_{k=1}^t vecho(\boldsymbol{x}_k \boldsymbol{x}'_k) - \frac{1}{T-t}\sum\limits_{k=t+1}^T vecho(\boldsymbol{x}_k \boldsymbol{x}'_k) \right),
		\end{aligned}
	\end{equation*}
	where $vecho(\cdot)$ indicates the half-vectorization $p(p-1)/2$ vector by vectorizing only the lower triangular part without the diagonal of the symmetric matrix, and $``\circ"$ is the Hadamard product. If a change point exists at $t_0$, then one can verify that $\boldsymbol{v}_{t_0}$ is an estimator of $vecho(R_1-R_2)^2$, which measures the difference between the two population correlation matrices. In particular, the expectation of the component $\boldsymbol{v}_t(i,j)$ of the vector $\boldsymbol{v}_t$ corresponding to the position $(i, j)$ in the matrices $R_1$ and $R_2$ is (for simplicity, this expectation is calculated under the assumption that original data $\{\boldsymbol{y}_t \}$ have mean zero and variance one. For notational convenience, write $\rho_1=\rho_1(i,j)$ and $\rho_2=\rho_2(i,j)$ here)
    \small	\begin{eqnarray}\label{exp_v}
    	&&E \boldsymbol{v}_t(i,j)\\ \nonumber
    	&=& \left\{ \begin{array}{lr}
    		(\rho_1-\rho_2)^2 + \frac{1}{t_0}\left(\beta_1 - \rho_1^2 \right) + \frac{1}{T-t_0} \left(\beta_2 - \rho_2^2 \right)  & \text{if} \ t=t_0,\\[1mm]
    		\frac{(T-t_0)^2}{(T-t)^2} (\rho_1-\rho_2)^2 + \left(\frac{1}{t} + \frac{t_0-t}{(T-t)^2} \right) \left(\beta_1 - \rho_1^2 \right) + \frac{T-t_0}{(T-t)^2} \left(\beta_2 - \rho_2^2 \right) & \text{if} \ t<t_0,\\[2mm]
    		\frac{t_0^2}{t^2} (\rho_1-\rho_2)^2 + \frac{t_0}{t^2}\left(\beta_1 - \rho_1^2 \right) + \left( \frac{t-t_0}{t^2} + \frac{1}{T-t} \right) \left(\beta_2 - \rho_2^2 \right) & \text{if} \ t>t_0.
    	\end{array} \right.
    \end{eqnarray}
where $\beta_1 := \mathbb{E}(x_{ki}x_{kj})^2$ for $k \le t$, $\beta_2 := \mathbb{E}(x_{ki}x_{kj})^2$ for $k > t$, and w.l.o.g, we can assume $\beta_1$ and $\beta_2$ are uniformly bounded in this paper (see Assumption ~\ref{A1}, Lemma~\ref{sub_exp}, and the Proposition 2.7.1 of \cite{vershynin2018}). Thus, in each of the three cases of (\ref{exp_v}), the first term is the main term with order $O(1)$, while the other terms are all $o(1)$.
Therefore, the expectation of $\boldsymbol{v}_t(i,j)$ consistently achieves the largest value at the true change point position $t_0$ because the coefficients $\frac{(T-t_0)^2}{(T-t)^2}$ and $\frac{t_0^2}{t^2}$ before $(\rho_1-\rho_2)^2$ are smaller than 1 when $t\neq t_0$. Moreover, these coefficients are unrelated to the position $(i,j)$. Therefore, for any fixed $t$, larger values of $\boldsymbol{v}_t(i,j)$ indicate a significant difference between $\rho_1(i,j)$ and $\rho_2(i,j)$. In contrast, if there is no change point in the sequence of correlation matrices, then all elements in the vectors $\boldsymbol{v}_t$'s $(1\leq t\leq T)$ are considerably small as $E \boldsymbol{v}_t(i,j)=o(1)$ for all $t$.
	
	In addition, instead of investigating all values of $t$ separately, we use a weighted summation
	\begin{equation}\label{W}
		\boldsymbol{w}=\frac{1}{T-3}\sum\limits_{t=2}^{T-2} \frac{t(T-t)}{T}\boldsymbol{v}_t 
	\end{equation}
	to identify the largest components among the $p(p-1)/2$ entries. The weights $\frac{t(T-t)}{T}$ are introduced to address the different sizes of the variance of $\boldsymbol{v}_t$ for different values of $t$. By examining the largest entries in the vector $\boldsymbol{w}$, we can detect whether change points exist in the correlation structure. 
	
	We propose a threshold based on signflip parallel analysis, and the Algorithm is as follows.
	
	$$
	\begin{array}{ll}
		\hline
		\multicolumn{2}{l}{{\bf Algorithm \ 1}: \text{Threshold via Signflip Permutation}}\\
		\hline
		\ & {\bf Input}:  \text{Data Matrix} \ Y \in \mathbb{R}^{p\times T}  (p \ \text{variables and} \ T \ \text{series}), \text{number of trials} \ q.\\
		\ & {\bf Output}: \text{Threshold} \ \tau_1.\\
		{\bf \scriptstyle 1} &{\bf for}\ m \leftarrow 1\ {\bf to}\ q \ {\bf do}\\
		{\bf \scriptstyle 2} &\hskip0.6cm \text{Randomly signflip entries of}\ Y: \text{form} \ \mathcal{R}_m\circ Y \ \text{where}\\
		& \hskip2cm \mathcal{R}_m(ij)\overset{i.i.d.}{\sim} \left\{\begin{array}{cc} +1, & \text{with probability 1/2},\\
			-1, & \text{with probability 1/2}, \end{array} \right.\\
		&\hskip0.7cm \text{that is} \ \mathcal{R}_m\in \mathbb{R}^{p\times T} \ \text{has independent identically distributed Rademacher entries};\\
		{\bf \scriptstyle 3}& \hskip0.6cm \text{Calculate vectors} \ \{\tilde{\boldsymbol{v}}_t^{(m)} \}_{2\leq t\leq T-2}  \ \text{defined in} \ (\ref{Vt});\\
		{\bf \scriptstyle 4}& \hskip0.6cm \text{Calculate the weighted sum} \ \tilde{\boldsymbol{w}}^{(m)} \ \text{in} \ (\ref{W}) \ \text{from} \ \{\tilde{\boldsymbol{v}}_t^{(m)} \}_{2\leq t\leq T-2}; \\
		{\bf \scriptstyle 5} & {\bf end}\\
		{\bf \scriptstyle 6} & \tau_1 \leftarrow \text{the largest element in} \ W=(\tilde{\boldsymbol{w}}^{(1)},\ldots,\tilde{\boldsymbol{w}}^{(q)}) \ \text{that is,} \ \tau_1=max(W).\\
		\hline
	\end{array}
	$$
	
	Let $\boldsymbol{w}(i, j)$ denote the elements of the vector $\boldsymbol{w}$ corresponding to the position $(i, j)$ in the matrices $R_1$ and $R_2$. We identify all components which are larger than the threshold $\tau_1$, to be precise, define
	\begin{eqnarray}
		\boldsymbol{w}_{\tau_1}=\{(i,j):\boldsymbol{w}(i,j)>\tau_1,1\le i< j\le p\},
	\end{eqnarray}
	as the index set of all corresponding components. Then, for the testing problem in (\ref{hypothesis_detection}), the rejection region is 
	\begin{eqnarray}\label{rejection_region}
		\mathcal{C}= \{\boldsymbol{w}: card(\boldsymbol{w}_{\tau_1})>0 \}  ,
	\end{eqnarray}
	where $card(\cdot)$ is the cardinality of a set. The change point detection approach based on the rejection region $\mathcal{C}$ in Equation (\ref{rejection_region}) will be referred to as the signflip parallel analysis-based change point detection (SPAD) procedure. 
	
	\begin{remark1}
		To illustrate why the threshold $\tau_1$ selected by signflip permutation is valid, we examine the expectation of the component $\tilde{\boldsymbol{v}}_t(i,j)$ of the vector $\tilde{\boldsymbol{v}}_t$ for signflipped data $\mathcal{R}_m\circ Y$ (under the same assumption as (\ref{exp_v})). We obtain
	    
	    		\begin{eqnarray}
	    	E \tilde{\boldsymbol{v}}_t (i,j) = \left\{ \begin{array}{ll}
	    		\frac{1}{t_0}\beta_1+\frac{1}{T-t_0}\beta_2 & t=t_0,\\[1mm]
	    	 \left(\frac{1}{t} + \frac{t_0-t}{(T-t)^2}  \right) \beta_1 + \frac{T-t_0}{(T-t)^2} \beta_2   & t<t_0,\\ [2mm]
	    	  \frac{t_0}{t^2}\beta_1 + \left( \frac{t-t_0}{t^2} + \frac{1}{T-t} \right) \beta_2 & t>t_0.
	    	\end{array} \right. 
	    \end{eqnarray}

Compared to the expectation of $\boldsymbol{v}_t(i,j)$ in (\ref{exp_v}), 
the elements in the vectors $\tilde{\boldsymbol{v}}_t$'s are considerably small for all $t$ both under the null and alternative hypothesis, as the expectation is not related to the difference between $\rho_1(i,j)$ and $\rho_2(i,j)$, and $E \tilde{\boldsymbol{v}}_t (i,j)=o(1)$. In addition, we illustrate the phenomenon in Figure \ref{fig_w} through a simulation study. Independent samples are drawn from Gaussian distribution with $p=100$ and $T=100$. Under $H_0$, the correlation matrices are $R_1=R_2=I_p$, and under $H_1$, $R_1=I_p$, and $R_2(i,j)=0.5 $  for all $1\le i\neq j \le p$. With increasing order, we plot the distributions of the elements of $\boldsymbol{w}$ for original data and the elements of $\tilde{\boldsymbol{w}}$ for signflipped data. In Figure \ref{fig_w}(a), it is obvious that the two distributions are almost coincident under $H_0$, except that the maximum value of $\tilde{\boldsymbol{w}}$ may be larger than that of $\boldsymbol{w}$. In contrast, under $H_1$, the distribution of the elements of $\boldsymbol{w}$ is always above that of $\tilde{\boldsymbol{w}}$, as shown in Figure \ref{fig_w}(b). More importantly,  the distributions of the elements of $\tilde{\boldsymbol{w}}$ are very similar under both $H_0$ and $H_1$. Thus, the signflip step breaks the changing structure in the original data. See \cite{hong2020selecting} for more signflip parallel analysis properties.
		\begin{figure} \centering
			\subfigure[]{\includegraphics[width=0.45\textwidth]{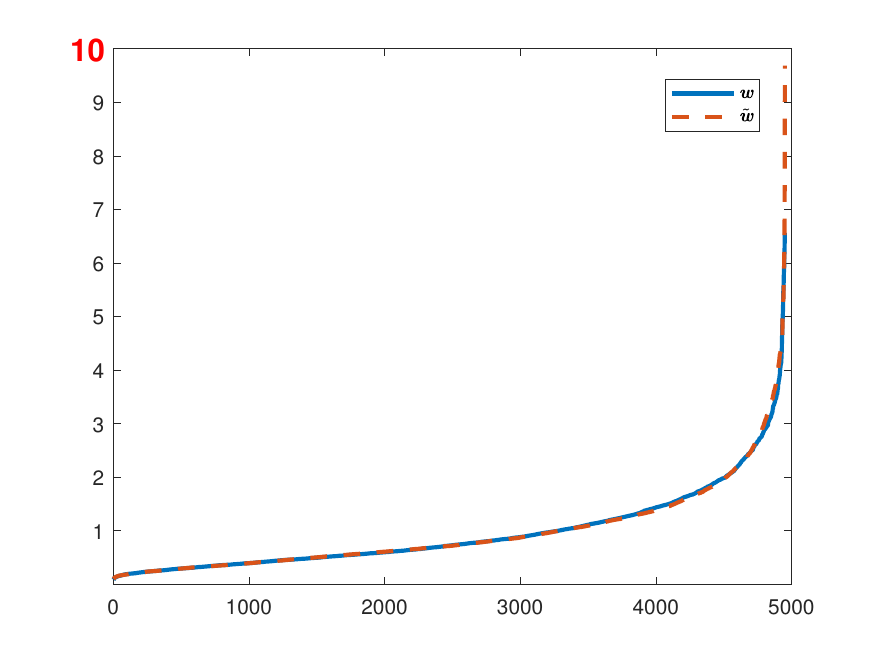}}
			\subfigure[]{\includegraphics[width=0.45\textwidth]{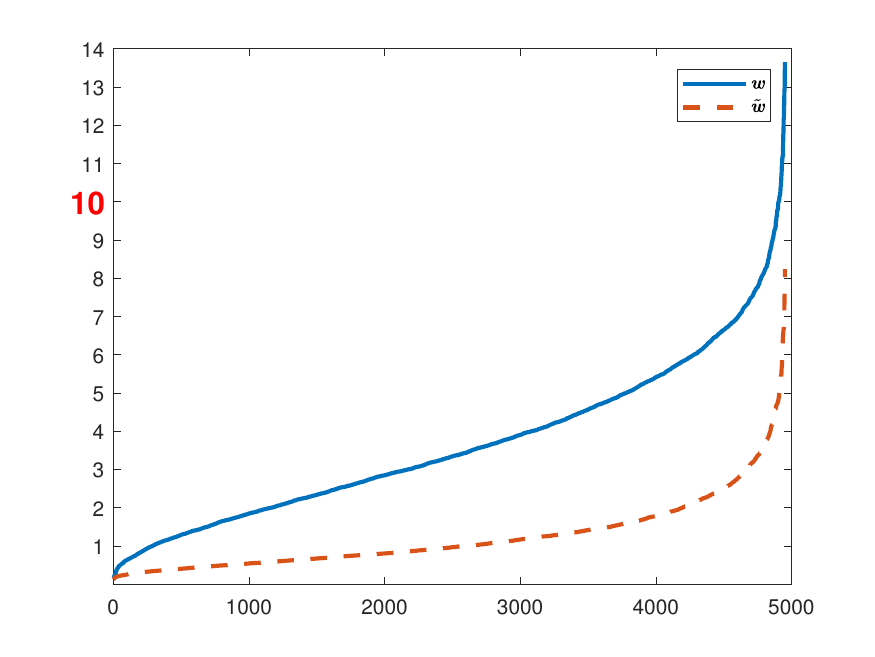}}
			\caption{The distributions of increasing-order elements of $\boldsymbol{w}$ and $\tilde{\boldsymbol{w}}$, (a) under $H_0$, (b) under $H_1$.}
			\label{fig_w}
		\end{figure}
	\end{remark1}

	
	\section{Estimation of Change Point Location}
	
	In many applications, if the null hypothesis $H_0: corr(\boldsymbol{y}_1)=\cdots=corr(\boldsymbol{y}_T)$ is rejected, it is often of significant interest to further estimate the change point's location and investigate at which components the correlation matrices $R_1$ and $R_2$ differ from each other, that is recovering the support of $R_1-R_2$. Suppose the change point location in the sequence is $t_0=\beta T, 0<\beta<1$, under the alternative with one change point, we aim to estimate the change point fraction $\beta$.
	
	When the null hypothesis $H_0$ in (\ref{hypothesis_detection}) is rejected, the cardinality of $\boldsymbol{w}_{\tau_1}$ is nonzero, and its element $\{\boldsymbol{w}_{\tau_1}(i,j)\}$ indicates a significant difference between $\rho_1(i,j)$ and $\rho_2(i,j)$. Thus, based on the components corresponding to $\boldsymbol{w}_{\tau_1}$, we can estimate the location of the change point. In order to keep more sufficiently significant changes, we use a new threshold $\tau_2$, which is slightly smaller than $\tau_1$. $\tau_2$ is generated the same way as $\tau_1$ by Algorithm 1 except for Line 6, $\tau_2$ is the $\alpha$-quantile $(95\%, 90\%)$ of all elements of $W$, that is,
	\begin{eqnarray}\label{tau2}
		\tau_2= \alpha \ \text{- quantile of} \ vec(W),
	\end{eqnarray}
	where $vec(\cdot)$ is the vectorization of a matrix. 
	Then, we reduce the $p(p-1)/2$-dimensional vector $\boldsymbol{w}$ to a vector of dimension $d$,
	\begin{eqnarray}\label{w_tau2}
		\boldsymbol{w}_{\tau_2}=\{(i,j):\boldsymbol{w}(i,j)>\tau_2,1\le i< j\le p\},
	\end{eqnarray}
	and correspondingly, we obtain $\boldsymbol{z}_i$, a $d$-dimensional subvector of vecho($\boldsymbol{x}_i \boldsymbol{x}'_i$) by keeping the $d$ components of the index set $\boldsymbol{w}_{\tau_2}$.
	
	For the estimation of the change point, we consider the CUSUM statistic
	\begin{equation}
		\begin{aligned}
			U_T(t)=\frac{1}{T^4} \sum_{i,k=1}^{t} \sum_{j,l=t+1}^T  (\boldsymbol{z}_i - \boldsymbol{z}_j)'(\boldsymbol{z}_k - \boldsymbol{z}_l)
			=\frac{1}{T^4}||(T-t)t (\tilde{\rho}_1^t - \tilde{\rho}_{t+1}^T)||_2^2,
		\end{aligned}
	\end{equation}
	where  $||\cdot||_2$ represents the Euclidean norm, and $\tilde{\rho}_{1}^t$ and $\tilde{\rho}_{t+1}^T$ represent the vectors containing the elements of the correlation matrice estimators $\hat{R}_{1}^t$ and $\hat{R}_{t+1}^T$ corresponding to positions identified in the dimension reduction index set $\boldsymbol{w}_{\tau_2}$, respectively. Then the estimator of the change point fraction $\beta$ is defined by
	\begin{eqnarray}\label{estimation}
		\hat{\beta}=\arg \max_{1\leq t\leq T} U_T(t)/T.
	\end{eqnarray}
	This estimator $\hat{\beta}$ will be referred to as the signflip parallel analysis-based CUSUM estimator (SPACE). The dimension reduction index set $\boldsymbol{w}_{\tau_2}$ denotes the support of $R_1-R_2$, that is, the positions at which the two correlation matrices differ.
	
	To illustrate the effectiveness of dimension reduction in (\ref{w_tau2}) and derive the asymptotic consistency of the proposed estimator $\hat{\beta}$ in (\ref{estimation}), we need the following assumptions on observations.
	\begin{assumption}\label{A1}
		Denote $\boldsymbol{x}_t=(x_{t1}, \ldots, x_{tp})',t=1,\ldots,T$. For any   $1 \le i \le p$, $x_{ti}$ is a sub-Gaussian random variable, that is, there are positive constants $C_1$, $C_2$ (independent of the indices $t$ and $i$) such that for every $\varepsilon>0$,
		$${P}(|x_{ti}>\varepsilon|)\le C_1e^{-C_2\varepsilon^2} .$$    
	\end{assumption}
	\begin{assumption}\label{A2}
		The smallest nonzero entry of the matrix $R_1-R_2$ satisfies
		\begin{eqnarray}
			\left|{\rho}_1(i,j)-{\rho}_2(i,j)\right| > C_3\sqrt{\frac{\tau_2}{T}}\cdot \max\left\{\frac{T^2}{(T-t_0)^2}, \frac{T^2}{t_0^2} \right\}, 1\le i\le j\le p, 
		\end{eqnarray}
		where $\tau_2$ is given in (\ref{tau2}).
	\end{assumption}
	\begin{assumption}\label{A3}
		For some positive constant c, we have
		\[p^2T^2e^{-c\sqrt{\tau_2}}=o(1).  \]
	\end{assumption}
	
	In the following theorem, we prove that all the entries with no difference are not included in $\boldsymbol{w}_{\tau_2}$.
	\begin{theorem2}\label{Theorem_1}
		Define $\mathcal{N}=\left\{(i,j):1\le i< j \le p; \rho_1 (i,j)=\rho_2(i,j)\right\}$ as the set of indices corresponding to identical elements in the correlation matrices $R_1$ and $R_2$. Then under Assumption \ref{A1}, 
		\begin{equation}
			\mathbb{P}\left\{\underset{(i,j)\in \mathcal{N}}{\bigcup}\left\{\boldsymbol{w}(i,j)>\tau_2\right\}\right\}=\mathbb{P}(\mathcal{N}\cap \boldsymbol{w}_{\tau_2} \neq \emptyset) \le  c_1p^2 T \left(e^{-c_2 T^{1/4}\sqrt{\tau_2}} + e^{-c_3 \min(T^{1/4}\sqrt{\tau_2},\tau_2)}  \right),
		\end{equation} 
		where $c_1$, $c_2$ and $c_3$ are some constants. In addition, under Assumption \ref{A3} we have 
		\begin{equation}
			\mathbb{P}\left\{\underset{(i,j)\in \mathcal{N}}{\bigcup}\left\{\boldsymbol{w}(i,j)>\tau_2\right\}\right\} \rightarrow 0.
		\end{equation}
	\end{theorem2}
	In the next theorem, we prove that all the entries with a difference larger than $C_3\sqrt{\frac{\tau_2}{T}}\cdot \max\left\{\frac{T^2}{(T-t_0)^2}, \frac{T^2}{t_0^2} \right\}$ are kept in $\boldsymbol{w}_{\tau_2}$.
	\begin{theorem2}\label{Theorem_2}
		Define $\mathcal{P}=\left\{(i,j):1\le i < j \le p; |\rho_1(i,j)-\rho_2(i,j)|>\lambda\right\}$ as the set of components which differ by more than $\lambda$, where $\lambda=C_3\sqrt{\frac{\tau_2}{T}}\cdot \max\left\{\frac{T^2}{(T-t_0)^2}, \frac{T^2}{t_0^2} \right\}$ as given in Assumption \ref{A2}. Then under Assumption \ref{A1} and Assumption \ref{A2}, we have 
		\begin{equation}
			\mathbb{P}\left\{\underset{(i,j)\in \mathcal{P}}{\bigcap}\{\boldsymbol{w}(i,j)>\tau_2\}\right\}
			=\mathbb{P}\left(\mathcal{P}\subset \boldsymbol{w}_{\tau_2}\right)\ge 1- c_4p^2 T e^{-c_5\sqrt{\tau_2T}} ,
		\end{equation}   
		where $c_4$ and $c_5$ are some constants. 
		In addition, under Assumption \ref{A3} we have
		\begin{equation}
			\mathbb{P}\left\{\underset{(i,j)\in \mathcal{P}}{\bigcap}\left\{\boldsymbol{w}(i,j)>\tau_2\right\}\right\} \rightarrow 1.
		\end{equation}
	\end{theorem2}
	Theorem~\ref{Theorem_2} shows that under Assumptions~\ref{A2} and \ref{A3}, the vector $\boldsymbol{w}_{\tau_2}$ recovers the true support of $R_1-R_2$ exactly with probability tending to 1. 
	
	In the next result, we establish the asymptotic consistency of the estimator $\hat{\beta}$. Here they symbol $\stackrel{p}{\rightarrow} $ denotes convergence in probability.
	\begin{theorem2}\label{theorem_3}
		Under Assumptions \ref{A1} and \ref{A2}, we have 
		$$\mathbb{P}\left\{ \left|\frac{\hat{\beta}}{\beta}-1\right|\ge \epsilon \right\} \le c_6p^2T^2e^{-c_7\sqrt{\tau_2}}, $$     
		where $c_6$ and $c_7$ are some constants. 
		In addition, under Assumption \ref{A3} we have
		$$\hat{\beta}\stackrel{p}{\rightarrow} \beta.$$
	\end{theorem2}
	
	\begin{remark1}
		The number of trails $q$ in the Algorithm. In a parallel analysis, the number of trials should be as large as possible to retain a stable and accurate result, such as 50 and 100 trials  \citep{horn1965rationale,humphreys1975investigation,turner1998effect}. However, many repetitions lead to time-consuming computation, especially for high-dimensional data. We investigate a suitable number of parallel trails by conducting simulations on the effect of the number of trials $q$ in both change point detection and estimation.  
		
		Figure~\ref{111} presents the success rate of the proposed detection method, SPAD, against the number of parallel trails $q$ under the $H_0$ without a change point. The success rate grows fast when $q$ is still small. Therefore, we recommend a number of trials of around 30 in the change point detection procedure.
		
		Figure~\ref{222} presents the mean values of  SPACE's estimated change point fraction against the number of parallel trails $q$ for the four cases in Section 5.2 and different combinations of $p$ and $T$. Each mean value is calculated from 200 repetitions. The mean values of estimates are all close to the true value and vary slightly with $q$. Therefore, the number of trails around 10 to 20 is enough to estimate the change point location.

		\begin{figure}[htbp]
			\centering
			\includegraphics[width=1\textwidth,scale=0.6]{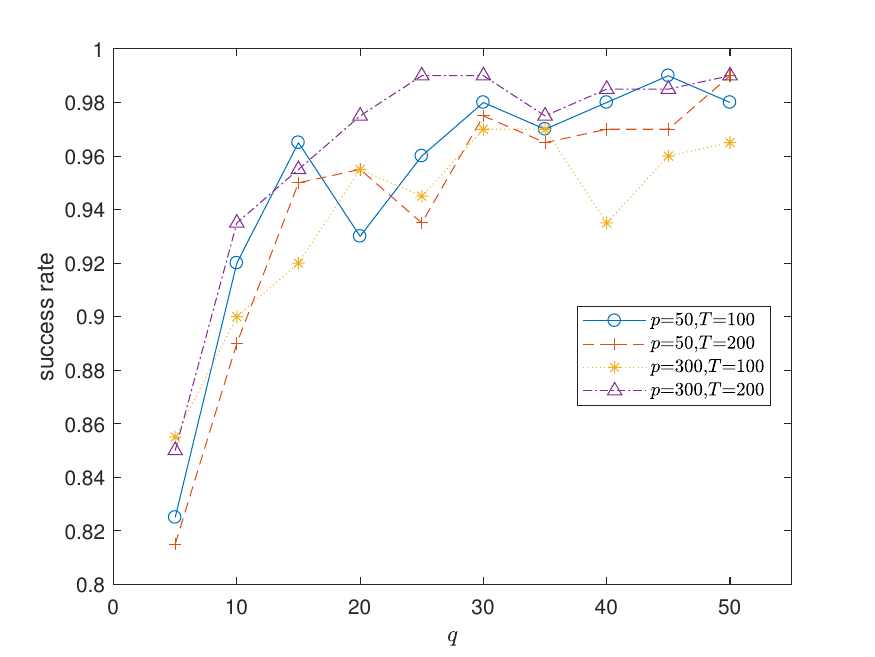}
			\caption{Success rate vs. the number of trails $q$ under $H_0$. }
			\label{111}
		\end{figure}

		\begin{figure}[htbp]
			\centering
			\includegraphics[width=1.2\textwidth,scale=1.5]{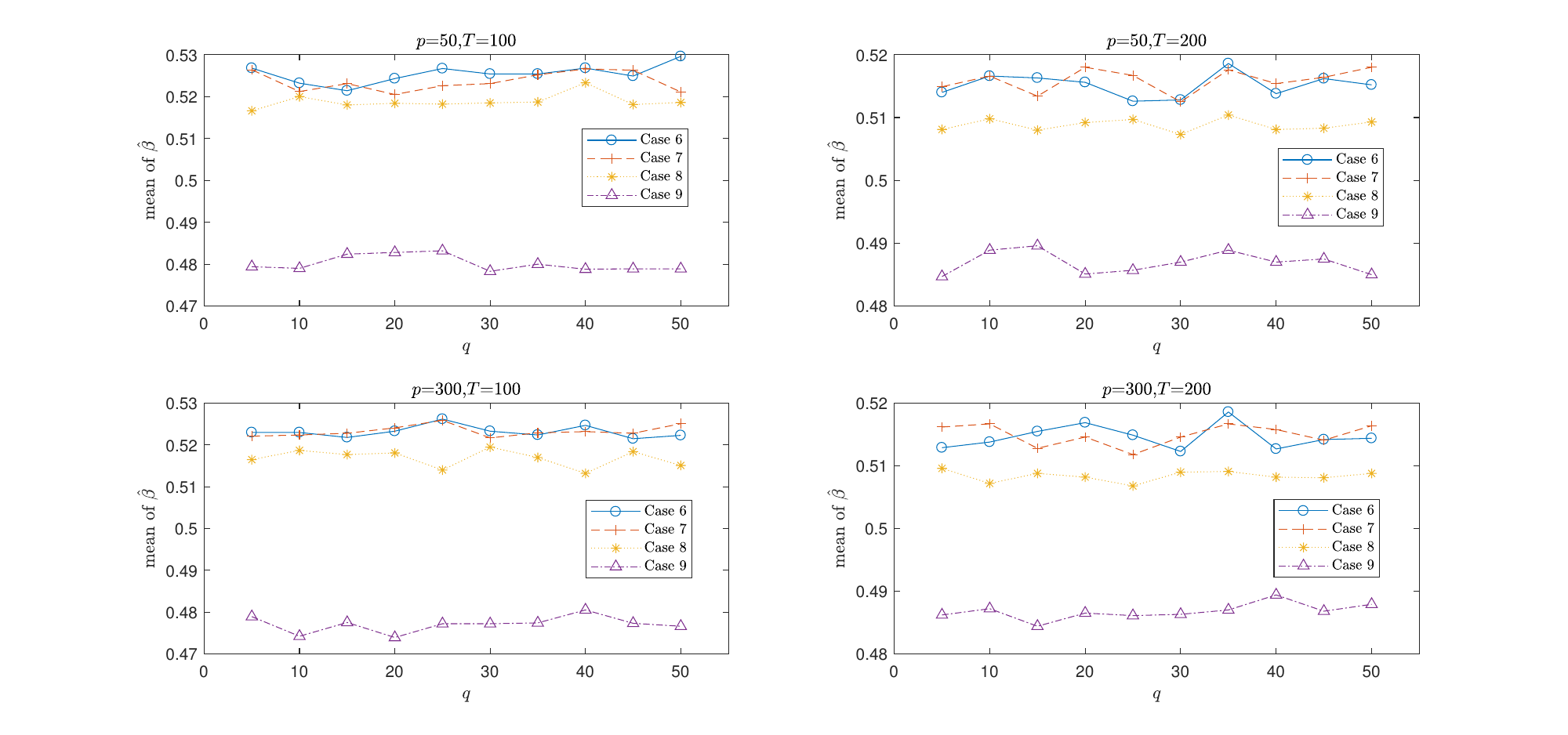}
			\caption{Mean value of the SPACE $\hat{\beta}$ vs. number of trails $q$.  }
			\label{222}
		\end{figure}
	\end{remark1}

	\section{Change Point in the Extreme Tail}
	Most existing literature aims to detect and estimate the change point in the middle of a data sequence. However, the change point hardly arises near the center and may appear in the tail. Therefore, in this section, we propose combining the SMOTE with SPACE to deal with the challenging situation where the change point arises in the extreme tail of a data sequence.
	
	SMOTE is a viral oversampling method proposed by \cite{chawla2002smote} and designed to deal with imbalanced data in classification problems by generating synthetic samples from the minority class. Specifically, for each sample $\boldsymbol{y}$ in the minority class, the nearest five neighbors with the smallest Euclidean distance are identified. One of them is randomly chosen as $\boldsymbol{y}_t^*$, based on which a new synthetic SMOTE sample is produced as $\boldsymbol{y}_s=\boldsymbol{y}_t+u \cdot (\boldsymbol{y}_t^* – \boldsymbol{y}_t )$, where $u$ is randomly chosen from a standard uniform distribution \citep{blagus2013smote}.
	
	To estimate the change point in the extreme tail effectively, the proposed SPACE can be enhanced by combining the SMOTE, and the Algorithm is as follows.
	
	$$
	\begin{array}{ll}
		\hline
		\multicolumn{2}{l}{{\bf Algorithm \ 2}: \text{SMOTE + SPACE}}\\
		\hline
		\ & {\bf Input}:  \text{Data Matrix} \ Y \in \mathbb{R}^{p\times T}  (p \ \text{variables and} \ T \ \text{series}), \text{small positive number} \ \epsilon.\\
		\ & {\bf Output}: \text{the estimator of the change point fraction} .\\
		{\bf  \scriptstyle 1} & \beta_0 \leftarrow \text{SPACE result from the original data.} \\
		{\bf  \scriptstyle 2} & \text{Let}\ [\gamma T+1,T] \ \text{be the minority class, get}\ (1-\gamma)T \ \text{SMOTE variables,} \gamma\geq 0.9.\\	
		{\bf  \scriptstyle 3} & \beta_1 \leftarrow \text{SPACE result from the inflated data.}\\
		{\bf \scriptstyle 4} & eps=abs(\beta_0-\beta_1). \\
		{\bf \scriptstyle 5} &{\bf while}\ eps>\epsilon  \ {\bf do}\\
		
		{\bf \scriptstyle 6}& \hskip0.6cm \beta_0 = \beta_1;\\
		{\bf \scriptstyle 7}& \hskip0.6cm \text{Repeat the above steps 2-4}; \\
		{\bf \scriptstyle 10} & {\bf end}\\
		{\bf \scriptstyle 11} & \beta_1 \leftarrow \text{the estimator of the change point fraction}.\\
		\hline
	\end{array}
	$$
	
	Similarly, by combining the SMOTE, the proposed SPAD method can also be enhanced to detect the existence of change points when they are in the extreme tail of the sequence of data, see Algorithm 3 in the Supplement Materials and corresponding simulations.

	\section{Finite Sample Properties}
	
	In this section, we conduct extensive simulations to examine the proposed SPAD method's success rate and the proposed SPACE's accuracy and compare them with several existing methods. We consider different dimensions $p$ ranging from 5 to 500 and sample sizes $T=100$ and $T=200$. All the numerical results below are calculated from 200 replications.
	
	\subsection{Detection of Change Point}
	We first illustrate the performance of the SPAD method under both $H_0$ and $H_1$. 
	
	Under $H_0$, independent samples are drawn from multivariate Gaussian distributions with correlation matrice $R_1=R_2=I_p$. The threshold $\tau_1$ is obtained from $q=30$ parallel trials. We compare SPAD with a method proposed by \cite{dette2020estimating}, which is designed to detect a change point in a sequence of covariance matrices. For the reader's convenience, we recall the method of \cite{dette2020estimating}.
	\begin{itemize}
		\item The detection method of \cite{dette2020estimating} is based on the following vector.
		\begin{eqnarray}\label{D_dettle}
			D&=&\frac{1}{T-3}\sum_{t=2}^{T-2}\frac{t(T-t)}{T} \bigg\{ \frac{1}{t(t-1)}\underset{i\neq j \le t}{\sum\sum}vech(\dot{\boldsymbol{y}}_i \dot{\boldsymbol{y}}'_i) \circ vech(\dot{\boldsymbol{y}}_j \dot{\boldsymbol{y}}'_j) \\ \nonumber
			&&+ \frac{1}{(T-t)(T-t-1)} \underset{i\neq j >t }{\sum \sum}vech(\dot{\boldsymbol{y}}_i \dot{\boldsymbol{y}}'_i) \circ vech(\dot{\boldsymbol{y}}_j \dot{\boldsymbol{y}}'_j)\\ \nonumber
			&& - \frac{2}{t(T-t)} \sum\limits_{i \le t} \sum\limits_{j>t} vech(\dot{\boldsymbol{y}}_i \dot{\boldsymbol{y}}'_i) \circ vech(\dot{\boldsymbol{y}}_j \dot{\boldsymbol{y}}'_j) \bigg\},
		\end{eqnarray}
		where $\dot{\boldsymbol{y}}_i=\boldsymbol{y}_i-\bar{\boldsymbol{y}}$ is the demeaned vector of data, and $vech(\cdot)$ indicates the half-vectorization $p(p+1)/2$ vector by vectorizing the lower triangular part of the symmetric matrix. With a threshold $\tau_0 \log (\log T)$, where $\tau_0$ is chosen by the bootstrap procedure stated in Section 2.3 of \cite{dette2020estimating}, no component in $D$ is larger than the threshold implies no change point exists in the sequence of covariance matrices.
	\end{itemize}
	\begin{table}[h]
		\centering 
		\caption{The success rate of detecting no change point under $H_0$.}
		\begin{tabular}{ccccccc}
			\hline
			$p$ & 20 & 50 & 100 & 200 & 300 & 500 \\ \hline
			SPAD & 0.980 & 0.970 & 0.985 & 0.985 & 0.970 & 0.990 \\
			Dette (2020) & 0.850 & 0.945 & 0.960 & 0.965 & 0.970 & 0.980 \\ \hline
		\end{tabular}
		\label{spad_h0}
	\end{table}
	Table 1 presents the success rate of detecting no change point. Both methods have very high accuracies, and SPAD performs slightly better than Dette's method.
	
	Under $H_1$, independent Gaussian samples are considered in five different settings:
	\begin{itemize}
		\item	Case 1: $\beta=0.5$, $R_1=I_p$, and $R_2(i,j)=0.5 $  for all $1\le i\neq j \le p$ ;
		\item	Case 2: $\beta=0.5$, $R_1=I_p$, and $R_2(i,j)=0.5 $  for all $1\le i\neq j \le [p/2]$, where $[p/2]$ is the integer part of $p/2$;
		\item	Case 3: $\beta=0.75$, $R_1=I_p$, and $R_2(i,j)=0.5 $  for all $1\le i\neq j \le p$ ;
		\item	Case 4: $\beta_1=1/3$, $\beta_2=2/3$,  $R_1=I_p$, and $R_2(i,j)=0.5 $  for all $1\le i\neq j \le p$, $R_3=R_1=I_p$; 
		\item	Case 5: $\beta_1=1/3$, $\beta_2=2/3$, $R_1=I_p$, $R_2(i,j)=0.5 $  for all $1\le i\neq j \le p$, and $R_3(i,j)=0.9 $  for all $1\le i\neq j \le p$; 
	\end{itemize}
	\begin{table}[h]
		\centering
		\caption{The success rate of detection change point under $H_1$.}
		\begin{tabular}{cccccccc}
			\hline
			& $p$ & 20 & 50 & 100 & 200 & 300 & 500 \\ \hline
			\multirow{2}{*}{Case 1} & SPAD & 0.770 & 0.860 & 0.810 & 0.865 & 0.910 & 0.895 \\
			& Dette (2020) & 0.745 & 0.555 & 0.540 & 0.415 & 0.360 & 0.225 \\ \hline
			\multirow{2}{*}{Case 2} & SPAD & 0.525 & 0.660 & 0.695 & 0.700 & 0.785 & 0.795 \\
			& Dette (2020) & 0.535 & 0.465 & 0.360 & 0.305 & 0.255 & 0.170 \\ \hline
			\multirow{2}{*}{Case 3} & SPAD & 0.495 & 0.560 & 0.500 & 0.575 & 0.635 & 0.670 \\
			& Dette (2020) & 0.530 & 0.365 & 0.280 & 0.250 & 0.215 & 0.115 \\ \hline
			\multirow{2}{*}{Case 4} & SPAD & 0.070 & 0.045 & 0.050 & 0.055 & 0.070 & 0.055 \\
			& Dette (2020) & 0.165 & 0.075 & 0.030 & 0.035 & 0.020 & 0.025 \\ \hline
			\multirow{2}{*}{Case 5} & SPAD & 0.965 & 0.955 & 0.980 & 0.980 & 0.990 & 0.980 \\
			& Dette (2020) & 0.875 & 0.770 & 0.715 & 0.625 & 0.560 & 0.455 \\ \hline
		\end{tabular}
		\label{spad_h1}
	\end{table}
	Table \ref{spad_h1} presents the success rate of detecting change points under the five cases. Overall, the SPAD method prominently outperforms Dette's method in all cases. Significantly, as the dimension $p$ becomes larger, the accuracy of SPAD increases while the accuracy of Dette's method decreases. When the change point exists in the middle of the sequence (Cases 1 \& 2), a larger magnitude of changes in the correlation matrices leads to higher detection accuracy, as expected. The detection accuracies of both methods are also affected by the location of the change point. Compared to Case 1, the change point location in Case 3 is close to the tail, so the detection accuracies of both methods are lower. When there is more than one change point in the sequence (Cases 4 \& 5), the performances of both methods are influenced by the change direction of correlations. In Case 4, the correlation matrix changes from an identity matrix to an equal correlation matrix and then changes back to the identity matrix. It is hard for both methods to detect the change points. In Case 5, the correlations increase from 0 to $0.5$ and then to $0.9$. Both methods can detect the existence of change points. Our SPAD method performs much better and reaches very high accuracies.

	\subsection{Estimation of Change Point}
	We investigate the finite sample properties of the proposed SPACE under different settings and compare it with two alternative methods proposed by \cite{dette2020estimating} and \cite{cabrieto2018capturing}, which are recalled as follows. For the threshold $\tau_2$ in SPACE, we set $\alpha= 95\%$ in Equation (\ref{tau2}).
	\begin{itemize}
		\item The estimator of \cite{dette2020estimating} is defined as follows.
		First, a critical value $\tau$ is constructed from the bootstrap approach stated in Section 2.3 of \cite{dette2020estimating}, based on which an index set is obtained from the vector $D$ in (\ref{D_dettle})
		\[\mathcal{D}_{\tau}=\{(i,j): D(i,j)>\tau, 1\leq i\leq j\leq p \}. \]
		Then Dette's estimator is 
		\begin{eqnarray}
			\hat{\beta}_{D}=\frac{1}{T}\underset{1\le t \le T}{\arg \max}\frac{1}{T^4} \underset{(i\neq s) = 1}{\sum\limits^{t}\sum\limits^{t}} \underset{(j\neq l) = t+1 }{\sum\limits^{T}\sum\limits^{T}} (\tilde{\boldsymbol{y}}_i - \tilde{\boldsymbol{y}}_j)^{'} (\tilde{\boldsymbol{y}}_s - \tilde{\boldsymbol{y}}_l),
		\end{eqnarray}
		where $\tilde{\boldsymbol{y}}_i$ is a subvector of $vech(\dot{\boldsymbol{y}}_i \dot{\boldsymbol{y}}'_i)$ by keeping the corresponding components of the index set $\mathcal{D}_{\tau}$.
		
		\item The estimator of \cite{cabrieto2018capturing} is kernel-based. The similarities of all pairs of $(\boldsymbol{y}_t,\boldsymbol{y}_{t'})$ are measured using Gaussian kernel,
		\[   g(\boldsymbol{y}_t,\boldsymbol{y}_{t'}) = \exp\left( \frac{-||\boldsymbol{y}_t - \boldsymbol{y}_{t'}||^2}{2h^2}  \right), t, t'=1,\ldots, T,\]
		where $h$ is the bandwidth parameter, equals the median Euclidean distance between all observations. Then the KCP-raw estimator is
		\begin{eqnarray}
			\hat{\beta}_{C} = \frac{1}{T} \underset{1\le t \le T}{\arg \min} \frac{1}{T}(v_{1,t} +v_{2,t}),
		\end{eqnarray}
		where $v_{1,t} = (t-1) - \frac{1}{t-1} \sum\limits_{i=2}^t \sum\limits_{j=2}^t g(\boldsymbol{y}_t,\boldsymbol{y}_{t'})$ measures the homogeneousness of phase before the time point $t$ and  $v_{2,t} = (T-t) - \frac{1}{T-t} \sum\limits_{i=t+1}^T \sum\limits_{j=t+1}^T g(\boldsymbol{y}_t,\boldsymbol{y}_{t'})$ measures the homogeneousness of phase after the time point $t$.
	\end{itemize}

	In the data generating process, $\{\boldsymbol{y}_t \}_{1\leq t\leq T}$ are generated from two distributions: (i) Multivariate normal, $N(\mathbf{0}, \Sigma_k), k=1,2$; and (ii) Student-t, $t_5$; with covariance matrix $\Sigma_k = \tilde{D}R_k\tilde{D}$, where $\tilde{D}=diag(\sigma_{1},\ldots,\sigma_{p})$, $R_1$ and $R_2$ are the correlation matrices before and after the change point, respectively. We consider homoscedastic senario so $\sigma_{j}=1, j=1,\ldots,p$. Four different choices for correlation matrices $R_1$ and $R_2$ are considered.
	\begin{itemize}
		\item Case 6: $R_1=I_p$, and $R_2(i,j)=0.5 $  for all $1\le i\neq j \le p$ ;
		\item Case 7: $R_1=I_p$, and $R_2(i,j)=0.5 $  for all $1\le i\neq j \le [p/2]$ ;	
		\item Case 8: $R_1=I_p$, and $R_2(i,j)=0.5 $  for all $1\le i\neq j \le [p/3]$,  $R_2(i,j)=0.2 $  for all $[p/3]+1 \le i\neq j \le [2p/3]$,  $R_2(i,j)=0.8 $  for all $[2p/3]+1 \le i\neq j \le p$ ;
		\item Case 9: $R_1(i,j)=0.5$ for all $1\le i\neq j \le p$, and $R_2(i,i+1)=R_2(i,i-1)=-0.5$ for all $1\le i \le p$.
	\end{itemize}
	\begin{table}[h]
		\centering
		\setlength\tabcolsep{4pt}
		\footnotesize
		\caption{Normal. Mean, standard deviation (SD), and mean squared error (MSE) of three estimators when the change point exists in the middle $\beta=0.5$, $T=100$.}
		\begin{threeparttable}[b]
			\begin{tabular}{ccccccccccccc}
				\hline
				\multirow{10}{*}{SPACE} &  & \multicolumn{5}{c}{Case 6} &  & \multicolumn{5}{c}{Case 7} \\ \cline{3-7} \cline{9-13} 
				& $p$ & 5 & 50 & 100 & 300 & 500 & $p$ & 5 & 50 & 100 & 300 & 500 \\
				& Mean & 0.5243 & 0.5231 & 0.5231 & 0.5248 & 0.5217 & Mean & 0.5163 & 0.5227 & 0.5236 & 0.5244 & 0.5201 \\
				& SD & 0.0671 & 0.0290 & 0.0327 & 0.0344 & 0.0312 & SD & 0.1136 & 0.0294 & 0.0326 & 0.0349 & 0.0288 \\
				& MSE & 0.0051 & 0.0014 & 0.0016 & 0.0018 & 0.0014 & MSE & 0.0131 & 0.0014 & 0.0016 & 0.0018 & 0.0012 \\ \cline{2-13} 
				&  & \multicolumn{5}{c}{Case 8} &  & \multicolumn{5}{c}{Case 9} \\ \cline{3-7} \cline{9-13} 
				& $p$ & 5 & 50 & 100 & 300 & 500 & $p$ & 5 & 50 & 100 & 300 & 500 \\
				& Mean & 0.5173 & 0.5165 & 0.5130 & 0.5187 & 0.5160 & Mean & 0.4869 & 0.4760 & 0.4759 & 0.4758 & 0.4759 \\
				& SD & 0.0469 & 0.0241 & 0.0223 & 0.0309 & 0.0249 & SD & 0.0316 & 0.0351 & 0.0365 & 0.0294 & 0.0337 \\
				& MSE & 0.0025 & 0.0008 & 0.0007 & 0.0013 & 0.0009 & MSE & 0.0012 & 0.0018 & 0.0019 & 0.0014 & 0.0017 \\ \hline
				\multirow{10}{*}{Dette's} &  & \multicolumn{5}{c}{Case 6} &  & \multicolumn{5}{c}{Case 7} \\ \cline{3-7} \cline{9-13} 
				& $p$ & 5 & 50 & 100 & 300 & 500 & $p$ & 5 & 50 & 100 & 300 & 500 \\
				& Mean & 0.5047 & 0.5059 & 0.5036 & 0.4948 & 0.4833 & Mean & 0.4152 & 0.4815 & 0.4786 & 0.4774 & 0.4480 \\
				& SD & 0.0959 & 0.0720 & 0.0646 & 0.1041 & 0.1202 & SD & 0.2186 & 0.1263 & 0.1189 & 0.1383 & 0.1574 \\
				& MSE & 0.0092 & 0.0052 & 0.0042 & 0.0108 & 0.0147 & MSE & 0.0548 & 0.0162 & 0.0145 & 0.0195 & 0.0273 \\ \cline{2-13} 
				&  & \multicolumn{5}{c}{Case 8} &  & \multicolumn{5}{c}{Case 9} \\ \cline{3-7} \cline{9-13} 
				& $p$ & 5 & 50 & 100 & 300 & 500 & $p$ & 5 & 50 & 100 & 300 & 500 \\
				& Mean & 0.5082 & 0.5101 & 0.5139 & 0.5140 & 0.5089 & Mean & 0.4870 & 0.4906 & 0.4909 & 0.4920 & 0.4945 \\
				& SD & 0.0735 & 0.0588 & 0.0536 & 0.0338 & 0.0700 & SD & 0.0350 & 0.0195 & 0.0188 & 0.0365 & 0.0146 \\
				& MSE & 0.0054 & 0.0035 & 0.0031 & 0.0013 & 0.0050 & MSE & 0.0014 & 0.0005 & 0.0004 & 0.0014 & 0.0002 \\ \hline
				\multirow{10}{*}{KCP-raw} &  & \multicolumn{5}{c}{Case 6} &  & \multicolumn{5}{c}{Case 7} \\ \cline{3-7} \cline{9-13} 
				& $p$ & 5 & 50 & 100 & 300 & 500 & $p$ & 5 & 50 & 100 & 300 & 500 \\
				& Mean & 0.5550 & 0.6379 & 0.6208 & 0.6208 & 0.6180 & Mean & 0.5581 & 0.6204 & 0.6647 & 0.6782 & 0.6696 \\
				& SD & 0.2790 & 0.1941 & 0.1597 & 0.1508 & 0.1554 & SD & 0.3367 & 0.2847 & 0.2280 & 0.1940 & 0.1934 \\
				& MSE & 0.0805 & 0.0565 & 0.0400 & 0.0372 & 0.0380 & MSE & 0.1162 & 0.0952 & 0.0788 & 0.0692 & 0.0660 \\ \cline{2-13} 
				&  & \multicolumn{5}{c}{Case 8} &  & \multicolumn{5}{c}{Case 9} \\ \cline{3-7} \cline{9-13} 
				& $p$ & 5 & 50 & 100 & 300 & 500 & $p$ & 5 & 50 & 100 & 300 & 500 \\
				& Mean & 0.5288 & 0.6227 & 0.6534 & 0.6828 & 0.6622 & Mean & 0.4845 & 0.3837 & 0.3737 & 0.3692 & 0.3629 \\
				& SD & 0.3028 & 0.2211 & 0.1945 & 0.1735 & 0.1714 & SD & 0.2030 & 0.1732 & 0.1642 & 0.1569 & 0.1686 \\
				& MSE & 0.0921 & 0.0637 & 0.0612 & 0.0634 & 0.0555 & MSE & 0.0412 & 0.0434 & 0.0428 & 0.0416 & 0.0471 \\ \hline
			\end{tabular}
		\end{threeparttable}
		\label{normal_middle}
	\end{table}

	\begin{table}[h]
		\centering
		\setlength\tabcolsep{4pt}
		\footnotesize
		\caption{ Student-t. Mean, standard deviation (SD), and mean squared error (MSE) of three estimators when the change point exists in the middle $\beta=0.5$, $T=100$.}
		\begin{threeparttable}[b]
			\begin{tabular}{ccccccccccccc}
				\hline
				\multirow{10}{*}{SPACE} &  & \multicolumn{5}{c}{Case 6} &  & \multicolumn{5}{c}{Case 7} \\ \cline{3-7} \cline{9-13} 
				& $p$ & 5 & 50 & 100 & 300 & 500 & $p$ & 5 & 50 & 100 & 300 & 500 \\
				& Mean & 0.5184 & 0.5459 & 0.5256 & 0.5312 & 0.5279 & Mean & 0.5028 & 0.5265 & 0.5221 & 0.5174 & 0.5223 \\
				& SD & 0.1311 & 0.1050 & 0.1100 & 0.1200 & 0.1129 & SD & 0.2041 & 0.1502 & 0.1413 & 0.1644 & 0.1472 \\
				& MSE & 0.0174 & 0.0131 & 0.0127 & 0.0153 & 0.0135 & MSE & 0.0413 & 0.0232 & 0.0204 & 0.0272 & 0.0220 \\ \cline{2-13} 
				&  & \multicolumn{5}{c}{Case 8} &  & \multicolumn{5}{c}{Case 9} \\ \cline{3-7} \cline{9-13} 
				& $p$ & 5 & 50 & 100 & 300 & 500 & $p$ & 5 & 50 & 100 & 300 & 500 \\
				& Mean & 0.5363 & 0.5343 & 0.5289 & 0.5287 & 0.5233 & Mean & 0.4936 & 0.4727 & 0.4673 & 0.4649 & 0.4680 \\
				& SD & 0.1253 & 0.1042 & 0.1119 & 0.1235 & 0.1054 & SD & 0.0749 & 0.0948 & 0.1030 & 0.1201 & 0.1279 \\
				& MSE & 0.0169 & 0.0120 & 0.0133 & 0.0160 & 0.0116 & MSE & 0.0056 & 0.0097 & 0.0116 & 0.0156 & 0.0173 \\ \hline
				\multirow{10}{*}{Dette's} &  & \multicolumn{5}{c}{Case 6} &  & \multicolumn{5}{c}{Case 7} \\ \cline{3-7} \cline{9-13} 
				& $p$ & 5 & 50 & 100 & 300 & 500 & $p$ & 5 & 50 & 100 & 300 & 500 \\
				& Mean & 0.4430 & 0.2622 & 0.1886 & 0.0909 & 0.0552 & Mean & 0.3172 & 0.1998 & 0.1324 & 0.0658 & 0.0434 \\
				& SD & 0.2088 & 0.2492 & 0.2398 & 0.1777 & 0.1293 & SD & 0.2686 & 0.2502 & 0.2121 & 0.1637 & 0.1114 \\
				& MSE & 0.0466 & 0.1184 & 0.1542 & 0.1987 & 0.2145 & MSE & 0.1052 & 0.1524 & 0.1799 & 0.2152 & 0.2209 \\ \cline{2-13} 
				&  & \multicolumn{5}{c}{Case 8} &  & \multicolumn{5}{c}{Case 9} \\ \cline{3-7} \cline{9-13} 
				& $p$ & 5 & 50 & 100 & 300 & 500 & $p$ & 5 & 50 & 100 & 300 & 500 \\
				& Mean & 0.4472 & 0.2999 & 0.1906 & 0.0866 & 0.0688 & Mean & 0.4852 & 0.3815 & 0.2891 & 0.1902 & 0.1070 \\
				& SD & 0.1955 & 0.2541 & 0.2412 & 0.1773 & 0.1554 & SD & 0.0825 & 0.2147 & 0.2471 & 0.2391 & 0.1887 \\
				& MSE & 0.0408 & 0.1042 & 0.1536 & 0.2022 & 0.2100 & MSE & 0.0070 & 0.0599 & 0.1052 & 0.1528 & 0.1899 \\ \hline
				\multirow{10}{*}{KCP-raw} &  & \multicolumn{5}{c}{Case 6} &  & \multicolumn{5}{c}{Case 7} \\ \cline{3-7} \cline{9-13} 
				& $p$ & 5 & 50 & 100 & 300 & 500 & $p$ & 5 & 50 & 100 & 300 & 500 \\
				& Mean & 0.5292 & 0.5582 & 0.5710 & 0.5812 & 0.5954 & Mean & 0.5460 & 0.5212 & 0.5111 & 0.5814 & 0.5691 \\
				& SD & 0.3020 & 0.2581 & 0.2512 & 0.2483 & 0.2495 & SD & 0.3411 & 0.3481 & 0.3467 & 0.3465 & 0.3516 \\
				& MSE & 0.0916 & 0.0696 & 0.0679 & 0.0679 & 0.0710 & MSE & 0.1179 & 0.1210 & 0.1197 & 0.1261 & 0.1278 \\ \cline{2-13} 
				&  & \multicolumn{5}{c}{Case 8} &  & \multicolumn{5}{c}{Case 9} \\ \cline{3-7} \cline{9-13} 
				& $p$ & 5 & 50 & 100 & 300 & 500 & $p$ & 5 & 50 & 100 & 300 & 500 \\
				& Mean & 0.5460 & 0.5372 & 0.5614 & 0.5951 & 0.5990 & Mean & 0.4992 & 0.4300 & 0.3961 & 0.4391 & 0.4222 \\
				& SD & 0.3089 & 0.3266 & 0.3221 & 0.3162 & 0.2988 & SD & 0.2183 & 0.2403 & 0.2400 & 0.2579 & 0.2627 \\
				& MSE & 0.0971 & 0.1075 & 0.1070 & 0.1085 & 0.0986 & MSE & 0.0474 & 0.0623 & 0.0681 & 0.0699 & 0.0747 \\ \hline
			\end{tabular}
		\end{threeparttable}
		\label{student_middle}
	\end{table}	
	
	\begin{table}[h]
		\centering
		\setlength\tabcolsep{4pt}
		\footnotesize
		\caption{ Normal. Mean, standard deviation (SD), and mean squared error (MSE) of three estimators when the change point fraction is $\beta=0.7$, T=100.}
		\begin{threeparttable}[b]
			\begin{tabular}{ccccccccccccc}
				\hline
				\multirow{10}{*}{SPACE} &  & \multicolumn{5}{c}{Case 6} &  & \multicolumn{5}{c}{Case 7} \\ \cline{3-7} \cline{9-13} 
				& $p$ & 5 & 50 & 100 & 300 & 500 & $p$ & 5 & 50 & 100 & 300 & 500 \\
				& Mean & 0.6315 & 0.6980 & 0.6959 & 0.7014 & 0.6987 & Mean & 0.5744 & 0.6763 & 0.6768 & 0.6861 & 0.6855 \\
				& SD & 0.1072 & 0.0311 & 0.0328 & 0.0265 & 0.0237 & SD & 0.1518 & 0.0530 & 0.0509 & 0.0446 & 0.0419 \\
				& MSE & 0.0161 & 0.0010 & 0.0011 & 0.0007 & 0.0006 & MSE & 0.0387 & 0.0034 & 0.0031 & 0.0022 & 0.0020 \\ \cline{2-13} 
				&  & \multicolumn{5}{c}{Case 8} &  & \multicolumn{5}{c}{Case 9} \\ \cline{3-7} \cline{9-13} 
				& $p$ & 5 & 50 & 100 & 300 & 500 & $p$ & 5 & 50 & 100 & 300 & 500 \\
				& Mean & 0.6690 & 0.7034 & 0.7039 & 0.7032 & 0.7060 & Mean & 0.6585 & 0.6410 & 0.6293 & 0.6223 & 0.6239 \\
				& SD & 0.0738 & 0.0168 & 0.0184 & 0.0115 & 0.0140 & SD & 0.0622 & 0.0693 & 0.0820 & 0.0922 & 0.0878 \\
				& MSE & 0.0064 & 0.0003 & 0.0004 & 0.0001 & 0.0002 & MSE & 0.0056 & 0.0083 & 0.0117 & 0.0145 & 0.0135 \\ \hline
				\multirow{10}{*}{Dette's} &  & \multicolumn{5}{c}{Case 6} &  & \multicolumn{5}{c}{Case 7} \\ \cline{3-7} \cline{9-13} 
				& $p$ & 5 & 50 & 100 & 300 & 500 & $p$ & 5 & 50 & 100 & 300 & 500 \\
				& Mean & 0.5908 & 0.5701 & 0.5777 & 0.5587 & 0.5359 & Mean & 0.4318 & 0.5098 & 0.5059 & 0.5338 & 0.4842 \\
				& SD & 0.1700 & 0.1743 & 0.1525 & 0.1753 & 0.1939 & SD & 0.2545 & 0.2369 & 0.2281 & 0.2116 & 0.2400 \\
				& MSE & 0.0407 & 0.0471 & 0.0381 & 0.0505 & 0.0643 & MSE & 0.1364 & 0.0920 & 0.0894 & 0.0722 & 0.1039 \\ \cline{2-13} 
				&  & \multicolumn{5}{c}{Case 8} &  & \multicolumn{5}{c}{Case 9} \\ \cline{3-7} \cline{9-13} 
				& $p$ & 5 & 50 & 100 & 300 & 500 & $p$ & 5 & 50 & 100 & 300 & 500 \\
				& Mean & 0.6309 & 0.6393 & 0.6221 & 0.6231 & 0.6082 & Mean & 0.6551 & 0.6633 & 0.6619 & 0.6611 & 0.6455 \\
				& SD & 0.1339 & 0.1304 & 0.1479 & 0.1250 & 0.1370 & SD & 0.0661 & 0.0567 & 0.0665 & 0.0711 & 0.1157 \\
				& MSE & 0.0226 & 0.0206 & 0.0278 & 0.0215 & 0.0271 & MSE & 0.0064 & 0.0045 & 0.0058 & 0.0065 & 0.0163 \\ \hline
				\multirow{10}{*}{KCP-raw} &  & \multicolumn{5}{c}{Case 6} &  & \multicolumn{5}{c}{Case 7} \\ \cline{3-7} \cline{9-13} 
				& $p$ & 5 & 50 & 100 & 300 & 500 & $p$ & 5 & 50 & 100 & 300 & 500 \\
				& Mean & 0.5971 & 0.7632 & 0.7800 & 0.7810 & 0.7881 & Mean & 0.5640 & 0.6664 & 0.6714 & 0.7835 & 0.7765 \\
				& SD & 0.3093 & 0.1665 & 0.1446 & 0.0944 & 0.0944 & SD & 0.3384 & 0.3006 & 0.2968 & 0.1801 & 0.1958 \\
				& MSE & 0.1057 & 0.0316 & 0.0272 & 0.0154 & 0.0166 & MSE & 0.1325 & 0.0910 & 0.0884 & 0.0393 & 0.0440 \\ \cline{2-13} 
				&  & \multicolumn{5}{c}{Case 8} &  & \multicolumn{5}{c}{Case 9} \\ \cline{3-7} \cline{9-13} 
				& $p$ & 5 & 50 & 100 & 300 & 500 & $p$ & 5 & 50 & 100 & 300 & 500 \\
				& Mean & 0.5794 & 0.6993 & 0.7458 & 0.8087 & 0.7920 & Mean & 0.6147 & 0.5434 & 0.5226 & 0.5147 & 0.5038 \\
				& SD & 0.3196 & 0.2409 & 0.2284 & 0.1191 & 0.1482 & SD & 0.2390 & 0.2441 & 0.2347 & 0.2303 & 0.2465 \\
				& MSE & 0.1162 & 0.0578 & 0.0540 & 0.0259 & 0.0303 & MSE & 0.0641 & 0.0838 & 0.0863 & 0.0871 & 0.0989 \\ \hline
			\end{tabular}
		\end{threeparttable}
		\label{normal_tail}
	\end{table}

	\begin{table}[h]
		\centering
		\setlength\tabcolsep{4pt}
		\footnotesize
		\caption{ Student-t. Mean, standard deviation (SD), and mean squared error (MSE) of three estimators when the change point fraction is $\beta=0.7$, $T=100$.}
		\begin{threeparttable}[b]
			\begin{tabular}{ccccccccccccc}
				\hline
				\multirow{10}{*}{SPACE} &  & \multicolumn{5}{c}{Case 6} &  & \multicolumn{5}{c}{Case 7} \\ \cline{3-7} \cline{9-13} 
				& $p$ & 5 & 50 & 100 & 300 & 500 & $p$ & 5 & 50 & 100 & 300 & 500 \\
				& Mean & 0.5954 & 0.6262 & 0.6235 & 0.6137 & 0.6363 & Mean & 0.5393 & 0.5846 & 0.5826 & 0.5602 & 0.5825 \\
				& SD & 0.1710 & 0.1598 & 0.1558 & 0.1778 & 0.1502 & SD & 0.2279 & 0.1902 & 0.1798 & 0.2120 & 0.2000 \\
				& MSE & 0.0400 & 0.0308 & 0.0300 & 0.0389 & 0.0265 & MSE & 0.0773 & 0.0493 & 0.0460 & 0.0643 & 0.0536 \\ \cline{2-13} 
				&  & \multicolumn{5}{c}{Case 8} &  & \multicolumn{5}{c}{Case 9} \\ \cline{3-7} \cline{9-13} 
				& $p$ & 5 & 50 & 100 & 300 & 500 & $p$ & 5 & 50 & 100 & 300 & 500 \\
				& Mean & 0.6353 & 0.6561 & 0.6295 & 0.6329 & 0.6314 & Mean & 0.6321 & 0.5763 & 0.5759 & 0.5628 & 0.5708 \\
				& SD & 0.1584 & 0.1421 & 0.1600 & 0.1755 & 0.1627 & SD & 0.1222 & 0.1593 & 0.1628 & 0.1687 & 0.1649 \\
				& MSE & 0.0292 & 0.0220 & 0.0304 & 0.0352 & 0.0311 & MSE & 0.0195 & 0.0406 & 0.0418 & 0.0472 & 0.0437 \\ \hline
				\multirow{10}{*}{\begin{tabular}[c]{@{}c@{}}Dette\\ (2020)\end{tabular}} &  & \multicolumn{5}{c}{Case 6} &  & \multicolumn{5}{c}{Case 7} \\ \cline{3-7} \cline{9-13} 
				& $p$ & 5 & 50 & 100 & 300 & 500 & $p$ & 5 & 50 & 100 & 300 & 500 \\
				& Mean & 0.4872 & 0.2358 & 0.1740 & 0.0703 & 0.0370 & Mean & 0.3236 & 0.1734 & 0.1200 & 0.0621 & 0.0418 \\
				& SD & 0.2634 & 0.2727 & 0.2600 & 0.1703 & 0.0953 & SD & 0.2819 & 0.2610 & 0.2186 & 0.1671 & 0.1207 \\
				& MSE & 0.1143 & 0.2894 & 0.3440 & 0.4254 & 0.4487 & MSE & 0.2208 & 0.3450 & 0.3839 & 0.4346 & 0.4478 \\ \cline{2-13} 
				&  & \multicolumn{5}{c}{Case 8} &  & \multicolumn{5}{c}{Case 9} \\ \cline{3-7} \cline{9-13} 
				& $p$ & 5 & 50 & 100 & 300 & 500 & $p$ & 5 & 50 & 100 & 300 & 500 \\
				& Mean & 0.5100 & 0.2893 & 0.1585 & 0.0781 & 0.0489 & Mean & 0.6032 & 0.4105 & 0.2841 & 0.1979 & 0.1137 \\
				& SD & 0.2680 & 0.3041 & 0.2568 & 0.1788 & 0.1335 & SD & 0.1604 & 0.3065 & 0.3098 & 0.2897 & 0.2256 \\
				& MSE & 0.1076 & 0.2607 & 0.3588 & 0.4185 & 0.4417 & MSE & 0.0349 & 0.1773 & 0.2684 & 0.3356 & 0.3943 \\ \hline
				\multirow{10}{*}{KCP-raw} &  & \multicolumn{5}{c}{Case 6} &  & \multicolumn{5}{c}{Case 7} \\ \cline{3-7} \cline{9-13} 
				& $p$ & 5 & 50 & 100 & 300 & 500 & $p$ & 5 & 50 & 100 & 300 & 500 \\
				& Mean & 0.5587 & 0.6408 & 0.6515 & 0.6966 & 0.6984 & Mean & 0.5359 & 0.5352 & 0.5593 & 0.6358 & 0.5956 \\
				& SD & 0.3283 & 0.2990 & 0.3048 & 0.2714 & 0.2647 & SD & 0.3454 & 0.3669 & 0.3730 & 0.3625 & 0.3699 \\
				& MSE & 0.1272 & 0.0925 & 0.0948 & 0.0733 & 0.0697 & MSE & 0.1456 & 0.1611 & 0.1583 & 0.1348 & 0.1470 \\ \cline{2-13} 
				&  & \multicolumn{5}{c}{Case 8} &  & \multicolumn{5}{c}{Case 9} \\ \cline{3-7} \cline{9-13} 
				& $p$ & 5 & 50 & 100 & 300 & 500 & $p$ & 5 & 50 & 100 & 300 & 500 \\
				& Mean & 0.5703 & 0.6012 & 0.6035 & 0.6464 & 0.6405 & Mean & 0.6158 & 0.5565 & 0.5083 & 0.5528 & 0.5074 \\
				& SD & 0.3236 & 0.3516 & 0.3554 & 0.3374 & 0.3463 & SD & 0.2619 & 0.2759 & 0.2896 & 0.2977 & 0.3035 \\
				& MSE & 0.1210 & 0.1327 & 0.1350 & 0.1161 & 0.1229 & MSE & 0.0754 & 0.0963 & 0.1202 & 0.1098 & 0.1287 \\ \hline
			\end{tabular}
		\end{threeparttable}
		\label{student_tail}
	\end{table}	
	Table \ref{normal_middle} shows the mean change point fraction, SD, and MSE of three estimators for all four cases when data are drawn from Gaussian distribution. The true change point exists in the middle of the sequence. The SPACE and Dette's estimators perform better than the KCP-raw estimator in all cases. For Cases 6 and 8, Dette's estimator is slightly better than SPACE in terms of mean, while SPACE is better than Dette's estimator in terms of SD and MSE, which implies SPACE is more stable. Moreover, SPACE performs the best in Case 7 and Dette’s estimator performs the best in Case 9. As expected, the KCP-raw estimator is only close to the true value when $p=5$, but performs poorly when $p$ is large. Meanwhile, the KCP-raw estimator has the largest SD and MSE.  
	
	Table~\ref{student_middle} shows the mean change point fraction, SD, and MSE of three estimators for all four cases when data are drawn from Student-t distribution. SPACE performs best in all cases. With the results in Table \ref{normal_middle} (where data are Gaussian), Table~\ref{student_middle} shows the robustness of SPACE concerning population distributions. In contrast, Dette's estimator is not robust to non-Gaussian data as the mean values are far from the true value except for the cases when $p=5$.
	
	Table \ref{normal_tail} and Table \ref{student_tail} present the mean change point fraction, SD, and MSE of three estimators under four cases for normal and student-t distributed samples, respectively, when the change point is close to the tail, $\beta=0.7$. For normal samples, the proposed SPACE performs the best for Cases 6, 7, and 8 and is very close to the true value with the smallest SD and MSE. Dette's estimator performs slightly better than SPACE only under Case 9. For student-t samples, SPACE is the best, while Dette's estimator performs the worst since the mean tends to 0 as the dimension increases.

	\subsection{Change Point in the Extreme Tail}
	To show the performance of the SPACE+SMOTE Algorithm, we explore Case 6 when the change point occurs in the extreme right tail, $\beta=0.8, 0.9, 0.95$. The mean change point fraction, SD, MSE, and average iterations of the algorithm are shown in Table~\ref{smote_estimate}.
	As the dimension $p$ increases from 5 to 500, or the sample size $T$ increases from 100 to 200, the mean values become very close to the true value, and SD and MSE become smaller as expected. The algorithm is time-efficient as small number of iterations is enough.

	\begin{table}[]
		\setlength\tabcolsep{3pt}
		\centering
		\caption{Normal. Mean, standard deviation (SD), mean squared error (MSE), and iteration time(ITER) of SPACE+SMOTE algorithm for Case 6 when the change point exists in the extreme tail, $\epsilon=10^{-3}$.}
		\footnotesize
		\begin{tabular}{ccccccclcccccc}
			\hline
			\multicolumn{1}{l}{} & \multicolumn{6}{c}{T=100} &  & \multicolumn{6}{c}{T=200} \\ \cline{2-7} \cline{9-14} 
			\multirow{5}{*}{$\beta=0.8$} & $p$ & 5 & 50 & 100 & 300 & 500 &  & $p$ & 5 & 50 & 100 & 300 & 500 \\
			& Mean & 0.7072 & 0.7904 & 0.7966 & 0.8021 & 0.8019 &  & Mean & 0.7739 & 0.8047 & 0.8041 & 0.8056 & 0.8058 \\
			& SD & 0.1473 & 0.0523 & 0.0447 & 0.0358 & 0.0304 &  & SD & 0.0758 & 0.0126 & 0.0072 & 0.0089 & 0.0097 \\
			& MSE & 0.0302 & 0.0028 & 0.0020 & 0.0013 & 0.0009 &  & MSE & 0.0064 & 0.0002 & 0.0001 & 0.0001 & 0.0001 \\
			& ITER & 0.9450 & 1.1650 & 1.4050 & 1.3050 & 1.2850 &  & ITER & 0.7300 & 0.6100 & 0.5450 & 0.4200 & 0.4555 \\ \cline{1-7} \cline{9-14} 
			\multirow{4}{*}{$\beta=0.9$} & Mean & 0.7287 & 0.8306 & 0.8364 & 0.8570 & 0.8617 &  & Mean & 0.8267 & 0.9021 & 0.9012 & 0.9027 & 0.9024 \\
			& SD & 0.1884 & 0.1138 & 0.1077 & 0.0867 & 0.0733 &  & SD & 0.1320 & 0.0099 & 0.0163 & 0.0053 & 0.0052 \\
			& MSE & 0.0646 & 0.0177 & 0.0156 & 0.0093 & 0.0068 &  & MSE & 0.0227 & 0.0001 & 0.0003 & 0.0000 & 0.0000 \\
			& ITER & 1.1350 & 2.0000 & 2.4100 & 2.5000 & 2.7900 &  & ITER & 1.3950 & 1.8350 & 1.7650 & 1.6200 & 1.6250 \\ \cline{1-7} \cline{9-14} 
			\multirow{4}{*}{$\beta=0.95$} & Mean & 0.6642 & 0.7656 & 0.7814 & 0.8231 & 0.8393 &  & Mean & 0.7464 & 0.8785 & 0.9113 & 0.9269 & 0.9284 \\
			& SD & 0.2286 & 0.1378 & 0.1260 & 0.0968 & 0.0986 &  & SD & 0.1889 & 0.0948 & 0.0710 & 0.0343 & 0.0255 \\
			& MSE & 0.1336 & 0.0529 & 0.0442 & 0.0254 & 0.0216 &  & MSE & 0.0769 & 0.0141 & 0.0065 & 0.0017 & 0.0011 \\
			& ITER & 0.9650 & 2.2200 & 2.7450 & 3.5500 & 4.2400 &  & ITER & 1.0200 & 2.9600 & 3.5200 & 4.0450 & 4.1250 \\ \hline
		\end{tabular}
		\label{smote_estimate}
	\end{table}

\begin{figure} \centering
		\subfigure[]{\includegraphics[width=0.45\textwidth]{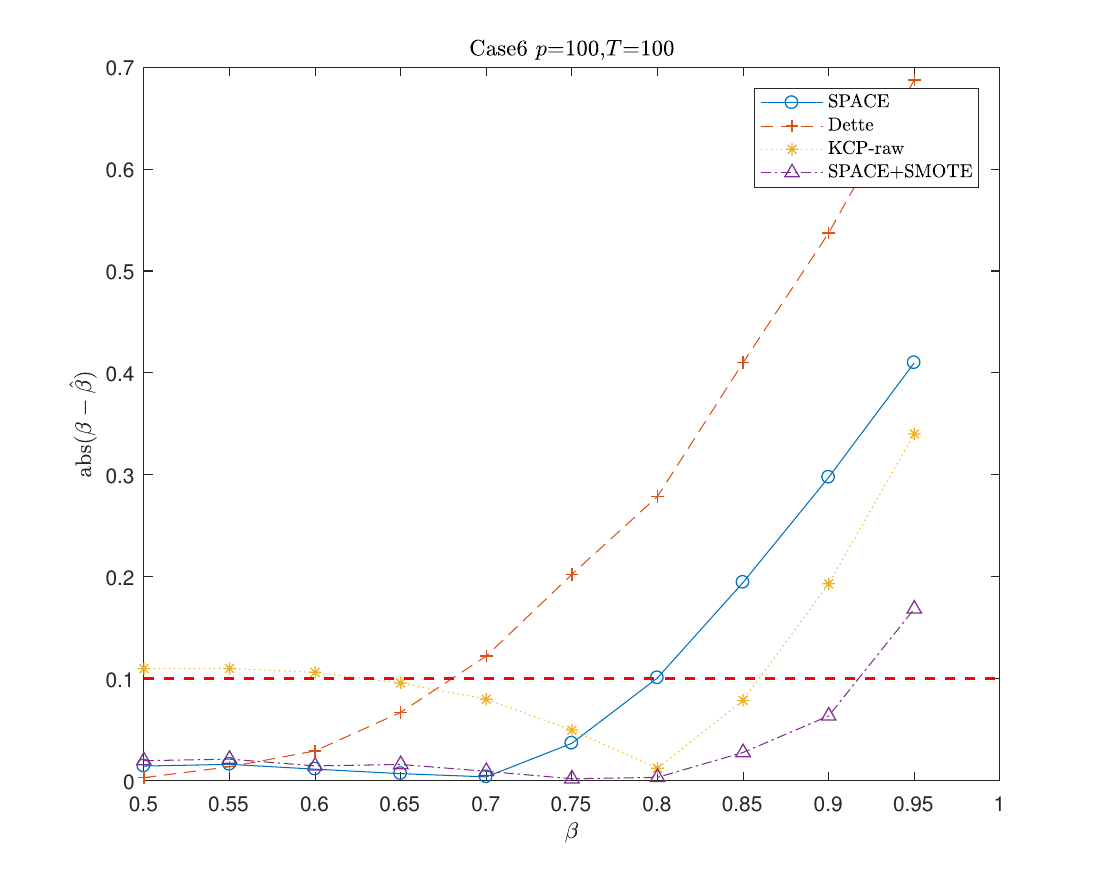}}
		\subfigure[]{\includegraphics[width=0.45\textwidth]{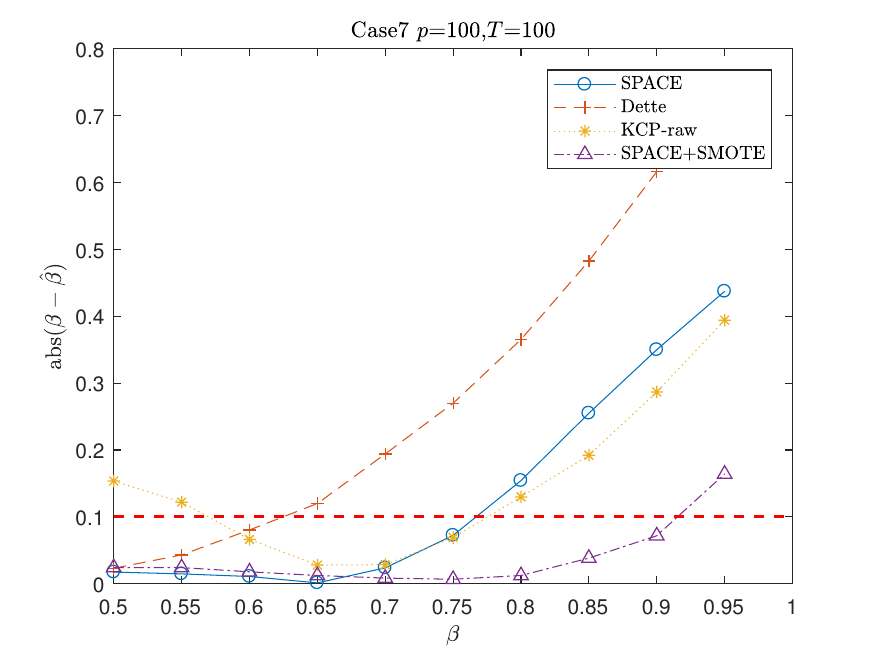}}
		\subfigure[]{\includegraphics[width=0.45\textwidth]{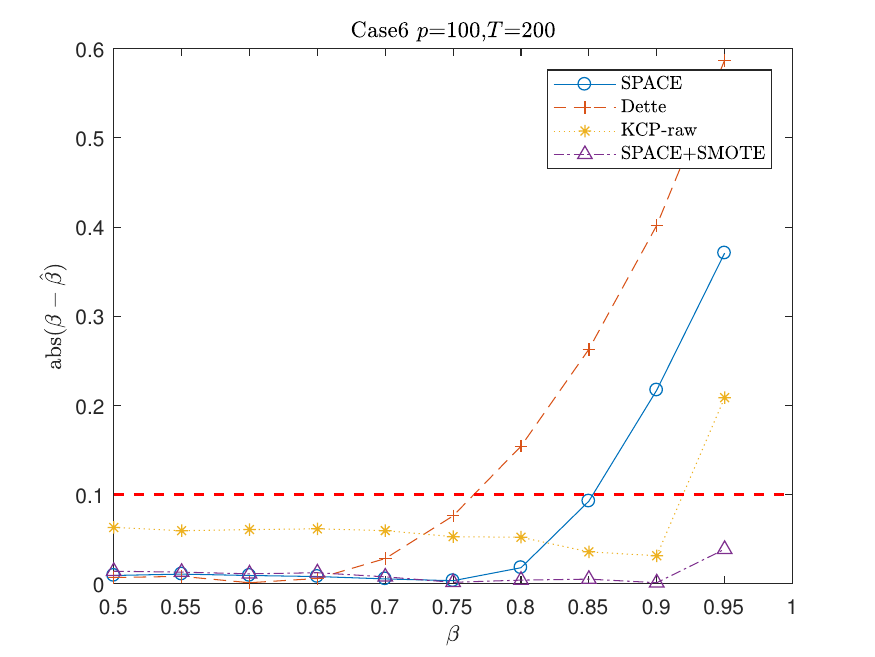}}
		\subfigure[]{\includegraphics[width=0.45\textwidth]{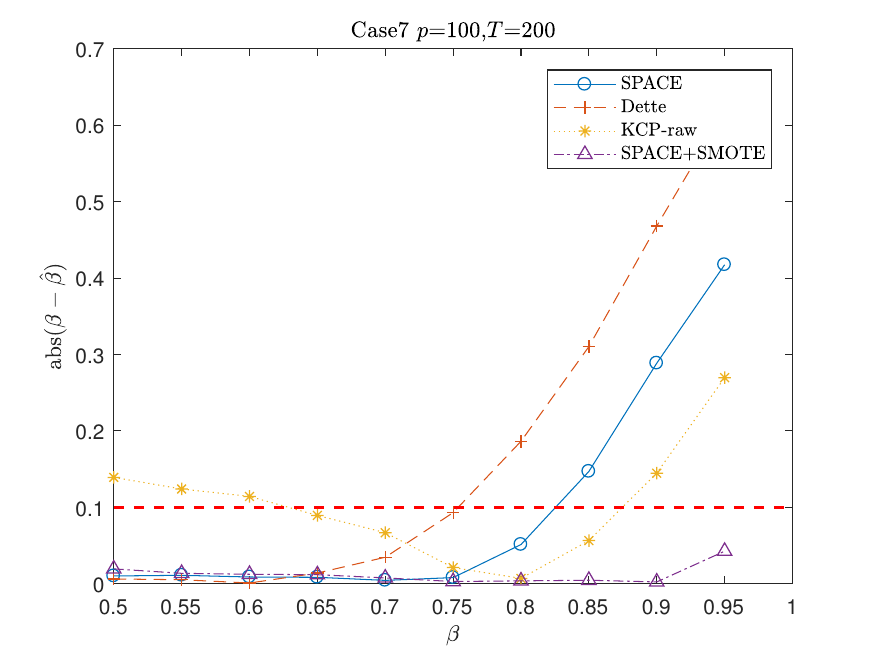}}
		\caption{The absolute values of $\beta-\hat{\beta}$ varies with $\beta$.}
		\label{fig_smote}
	\end{figure}

	To compare the overall performance of SPACE+SMOTE, SPACE, Dette's estimator and KCP-raw, Figure~\ref{fig_smote} shows the variation of the absolute value of $\hat{\beta}-\beta$ of these four methods as the true change point fraction $\beta$ changes from 0.5 to 0.95. A smaller absolute value indicates a more accurate estimated result. We consider Case 6 and Case 7 with $p=100, T=100, 200$ for normal samples. The SPACE+SMOTE Algorithm is the most effective one in estimating the location of the change point in the extreme tail. SPACE is also better than Dette (2020) and KCP-raw methods.

	\subsection{Heteroscedasticity and Dependence}
	Till now, homoscedasticity of random variables is assumed, which is often challenged in real data application, however. Therefore, we consider the following heteroscedasticity setting.
	$$\tilde{D} = diag(\sigma_{1},\ldots,\sigma_{p}),
	\quad \Sigma_1 = \tilde{D} R_1 \tilde{D},
	\quad \Sigma_2 = \tilde{D}  R_2  \tilde{D},$$
	where $\sigma_{i}, i=1,\ldots,p$ are independently drawn from Uniform distribution $U(1,21)$. We compare SPACE with Dette's estimator and KCP-raw under Cases 6-9 for normal and student-t distributed data, respectively. The results are similar to that for the homoscedasticity setting in the Section 5.2. The tables are in the supplementary materials.
	
	The independence assumption on the sequence of data $\boldsymbol{y}_1,\ldots, \boldsymbol{y}_T$ is also not common in real applications.
	We investigate the performance of SPACE for dependent data. Moreover, heteroscedastic scenarios are also considered. Following \cite{choi2020self}, we consider a vector autoregression of order 1,
	$$ 
	\boldsymbol{y}_t = \psi \boldsymbol{y}_{t-1} +e_t,
	$$
	where the error vector $\boldsymbol{e}_t = (e_{t1}, \ldots, e_{tp} )^{'} $ has zero mean and covariance matrix $\Sigma_t = D_t \tilde{R}_t D_t$ with $D_t = diag(\sigma_{t1}, \ldots, \sigma_{tp})$ and correlation matrix at time $t$, $\tilde{R}_t$. Here we consider the setting of correlation matrices in Case 6, that is, $\tilde{R}_t = R_1$ for $1\le t \le t_0$, $\tilde{R}_t = R_2$ for $t_0+1 \le t \le T$. The following four scenarios are considered.
	\begin{itemize}
		\item {\it Homoscedasticity serial dependence}: serial dependence is considered by $\psi = 0.8$, $\boldsymbol{e}_t = \Sigma_t^{1/2} \boldsymbol{\eta}_t$, $\sigma_{1i}=\cdots=\sigma_{pi} =1$ and $\boldsymbol{\eta}_t = (\eta_{t1}, \ldots, \eta_{tp})' $ is a vector of standard normal random variables. 
		
		\item {\it Heteroscedasticity serial dependence}: $\psi = 0.8$, $\boldsymbol{e}_t = \Sigma_t^{1/2} \boldsymbol{\eta}_t$, $\sigma_{1i}=\cdots=\sigma_{pi}1$ are independently from $U(1,21)$.
		
		\item {\it Conditional Heteroscedasticity}:  $\psi=0$, $\boldsymbol{e}_t = \Sigma_t^{1/2} \boldsymbol{\eta}_t$, $\sigma_{li}^2 = (1 - \alpha_1 -\alpha_2) + \alpha_1 a_{l,i-1}^2 +\alpha_2 \sigma_{l,i-1}^2$, $l=1,\ldots,p$, $(\alpha_1,\alpha_2) = (0.1,0.89)$.
		
		\item {\it  Unconditional Heteroscedasticity}:  $\psi=0$, $\boldsymbol{e}_t = \Sigma_t^{1/2} \boldsymbol{\eta}_t$, $\sigma_{li} = \sigma_0(1+ \delta \textit{I}(i> T/2)) $, $\sigma_0 = \delta =1$ for all $1\le l \le p$. 
	\end{itemize}
	
	The simulation results are presented in Table \ref{dependent}. Under both homoscedasticity serial dependence and conditional heteroscedasticity scenarios, SPACE and Dette's estimator perform better than the KCP estimator. Dette's estimator is better than SPACE in terms of mean, while SPACE is better than Dette's estimator in terms of SD and MSE. Under the unconditional heteroscedasticity scenario, the KCP estimator performs the best, and SPACE performs similarly with Dette's estimator.
	\begin{table}[h]
		\centering
		\caption{Mean, standard deviation (SD), and mean squared error (MSE) of three estimators under dependence and heteroscedasticity, $\beta=0.5$.}
		\setlength\tabcolsep{4pt}
		\begin{threeparttable}[b]
			\begin{tabular}{cccccccc}
				\hline
				\multirow{10}{*}{\begin{tabular}[c]{@{}c@{}}Homoscedasticity\\ Serial\\ Dependence\end{tabular}} & \multirow{4}{*}{SPACE} & $p$ & 5 & 50 & 100 & 300 & 500 \\ \cline{3-8} 
				&  & Mean & 0.5203 & 0.5503 & 0.5530 & 0.5562 & 0.5594 \\
				&  & SD & 0.1136 & 0.0769 & 0.0701 & 0.0655 & 0.0739 \\
				&  & MSE & 0.0133 & 0.0084 & 0.0077 & 0.0074 & 0.0090 \\ \cline{2-8} 
				& \multirow{3}{*}{Dette's} & Mean & 0.5093 & 0.5480 & 0.5455 & 0.5494 & 0.5566 \\
				&  & SD & 0.1387 & 0.0883 & 0.0748 & 0.0685 & 0.0818 \\
				&  & MSE & 0.0192 & 0.0101 & 0.0076 & 0.0071 & 0.0099 \\ \cline{2-8} 
				& \multirow{3}{*}{KCP-raw}& Mean & 0.5180 & 0.5601 & 0.5757 & 0.6306 & 0.6048 \\
				&  & SD & 0.2358 & 0.2055 & 0.1731 & 0.1700 & 0.1633 \\
				&  & MSE & 0.0557 & 0.0456 & 0.0355 & 0.0458 & 0.0375 \\ \hline
				\multirow{9}{*}{\begin{tabular}[c]{@{}c@{}}Heteroscedasticity\\ Serial\\ Dependence\end{tabular}} &
				\multirow{3}{*}{SPACE} & Mean & 0.5226 & 0.5381 & 0.5436 & 0.5434 & 0.5454 \\
				&  & SD & 0.1049 & 0.0536 & 0.0611 & 0.0453 & 0.0592 \\
				&  & MSE & 0.0115 & 0.0043 & 0.0056 & 0.0039 & 0.0055 \\ \cline{2-8} 
				& \multirow{3}{*}{Dette (2020)} & Mean & 0.5183 & 0.5357 & 0.5392 & 0.5379 & 0.5370 \\
				&  & SD & 0.1378 & 0.0668 & 0.0665 & 0.0528 & 0.0539 \\
				&  & MSE & 0.0192 & 0.0057 & 0.0059 & 0.0042 & 0.0043 \\ \cline{2-8} 
				& \multirow{3}{*}{KCP-raw} & Mean & 0.5573 & 0.6021 & 0.6050 & 0.6344 & 0.6252 \\
				&  & SD & 0.2604 & 0.2152 & 0.1958 & 0.1837 & 0.1664 \\
				&  & MSE & 0.0707 & 0.0565 & 0.0492 & 0.0516 & 0.0432 \\ \hline
				\multirow{9}{*}{\begin{tabular}[c]{@{}c@{}}Conditional\\ Heteroscedasticity\end{tabular}} & \multirow{3}{*}{SPACE} & Mean & 0.5135 & 0.5269 & 0.5242 & 0.5287 & 0.5295 \\
				&  & SD & 0.0894 & 0.0413 & 0.0401 & 0.0435 & 0.0419 \\
				&  & MSE & 0.0081 & 0.0024 & 0.0022 & 0.0027 & 0.0026 \\ \cline{2-8} 
				& \multirow{3}{*}{Dette's} & Mean & 0.5008 & 0.5227 & 0.5269 & 0.5272 & 0.5341 \\
				&  & SD & 0.1267 & 0.0864 & 0.0886 & 0.0846 & 0.0841 \\
				&  & MSE & 0.0160 & 0.0079 & 0.0085 & 0.0079 & 0.0082 \\ \cline{2-8} 
				& \multirow{3}{*}{KCP-raw} & Mean & 0.5137 & 0.6094 & 0.6219 & 0.6166 & 0.6016 \\
				&  & SD & 0.2627 & 0.1767 & 0.1683 & 0.1541 & 0.1477 \\
				&  & MSE & 0.0689 & 0.0430 & 0.0430 & 0.0372 & 0.0320 \\ \hline
				\multirow{9}{*}{\begin{tabular}[c]{@{}c@{}}Unconditional\\ Heteroscedasticity\end{tabular}} & \multirow{3}{*}{SPACE} & Mean & 0.5372 & 0.5316 & 0.5316 & 0.5292 & 0.5332 \\
				&  & SD & 0.0462 & 0.0414 & 0.0416 & 0.0347 & 0.0473 \\
				&  & MSE & 0.0035 & 0.0027 & 0.0027 & 0.0020 & 0.0033 \\ \cline{2-8} 
				& \multirow{3}{*}{Dette's}  & Mean & 0.5270 & 0.5300 & 0.5326 & 0.5349 & 0.5343 \\
				&  & SD & 0.0372 & 0.0400 & 0.0439 & 0.0544 & 0.0795 \\
				&  & MSE & 0.0021 & 0.0025 & 0.0030 & 0.0042 & 0.0075 \\ \cline{2-8} 
				& \multirow{3}{*}{KCP-raw} & Mean & 0.5297 & 0.5149 & 0.5139 & 0.5113 & 0.5130 \\
				&  & SD & 0.0622 & 0.0267 & 0.0266 & 0.0318 & 0.0257 \\
				&  & MSE & 0.0047 & 0.0009 & 0.0009 & 0.0011 & 0.0008 \\ \hline
			\end{tabular}
		\end{threeparttable}		
		\label{dependent}
	\end{table}

	\section{Real Data Analysis}
	Though the proposed SPACE outperforms the existing ones in the simulation experiments, we now compare SPACE and Dette's estimator on two real data sets. The KCP-raw method is omitted here due to its bad performance in the simulations. 
	\subsection{Gene Expression Profile}
	The first data is a real gene expression profile studied by \cite{sinnaeve2009gene}. The data set investigated the relationship between peripheral blood gene expression patterns and coronary artery disease severity. Patients undergoing coronary angiography were selected according to their coronary artery disease index (CADi), a validated angiographical measure of the extent of coronary atherosclerosis. RNA was extracted from the blood of 110 patients with at least a stenosis greater than $50 \%$ (CADi $\geq 23$ ) and from 112 controls without evidence of coronary stenosis (CADi$=0)$. Therefore, the total sample size is $n=222$. Rearrange the data so the true change point position is $t_0=110$. The raw dataset is available at the Gene Expression Omnibus repository with the accession number {\it Series GSE12288}.
	
	Following \cite{dette2020estimating}, we first apply the two sample $t$-tests with a significance level $0.05$ to all genes, and 1410 genes are reserved. As mentioned in \cite{sinnaeve2009gene}, many genes are not correlated with CADi, and thus it is suitable to do such a preprocessing step.
	
	To show the estimation performance of these two estimators, we change the number of genes kept in estimation procedures from $p=5$ to $p=1410$ corresponding to the smallest p-values in the two sample $t$-tests. Table \ref{real_data} records the estimated change point fractions for different $p$. As $p$ becomes large, both estimators are closer to the true value. SPACE performs better and remains stable when $p$ is large enough. In particular, SPACE achieves the true value when $p\geq 200$, while Dette's estimator never reaches the true value.
	\begin{table}[h]
		\centering
		\caption{ The true change point fraction is $\beta=110/222=0.4955$.}
		\setlength\tabcolsep{4pt}
		\small
		\begin{tabular}{cccccccccc}
			\hline
			$p$ & 5 & 20 & 50 & 100 & 200 & 300 & 500 & 1000 & 1410 \\ \hline
			SPACE & 0.4775 & 0.3694 & 0.4820 & 0.4910 & 0.4955 & 0.4955 & 0.4955 & 0.4955 & 0.4955 \\
			Dette's & 0.4730 & 0.4324 & 0.2342 & 0.2342 & 0.4279 & 0.4279 & 0.4279 & 0.2477 & 0.2477 \\ \hline
		\end{tabular}
		\label{real_data}
	\end{table}
	

	\subsection{FRED-MD Data Set}
	The second data set is the FRED-MD, a macroeconomic database of 134 monthly U.S.
	indicators from eight aspects (output and income, labor market, consumption and orders, orders
	and inventories, money and credit, interest rate and exchange rates, prices
	and stock market) from January 1959, aim to establish a convenient starting point for empirical analysis that requires "big data" \citep{mccracken2016fred}. The data set is updated in a timely manner and can be downloaded for free from the
	website {\em http://research.stlouisfed.org/econ/mccracken/sel/}.
	\cite{fan2020factor} used this data set to predict the risk of U.S. bond risk premia. As the change of correlation coefficient is often accompanied by risk, we apply the proposed method to estimate the change point location in this data set.
	
	After removing the missing data, 127 macroeconomic
	indicators are retained for nearly 20 years, that is, $p=127$. The latest month is July 2022. To show the estimation performance of these two estimators, we change the number of months $T$ until July 2022 from 50 to 200. Table~\ref{FRED-MD} records the estimated change point location (the exact month) for different $T$, with change point fraction in parentheses.
	
	\begin{table}[h]
		\centering 
	    \caption{The estimated change point month in the FRED-MD data set.}
		\begin{tabular}{cccccc}
			\hline
			$T$ & 50 & 100 & 130 & 150 & 200 \\
			start month & June 2018 & April 2014 & October 2011 & February 2010 & December 2005\\ \hline
			SPACE & Oct. 2021 & Apr. 2020 & Apr. 2020 & Apr. 2020 & Sep. 2008 \\
			& (0.82) & (0.73) &(0.79) & (0.82) &(0.17)\\
			Dette's & Feb. 2021 & Jan. 2021 & Oct. 2020 & Apr. 2020 & Apr. 2020 \\
			& (0.66) &(0.82) & (0.84) & (0.82) &(0.87)\\ \hline
		\end{tabular}
		\label{FRED-MD}
	\end{table}
	
	When $T=50$, the period is from June 2018 to July 2022. The SPACE estimated change point month is October 2021. This coincided with a sharp drop in U.S. stocks at the end of the third quarter of 2021, with the S\&P 500 posting its worst monthly drop since the coronavirus outbreak in March 2020, and the risk of stagflation loomed in the US economy, according to the reports in The Wall Street Journal and the BBC.
	
	When $T=100, 130, 150$, SPACE has been estimating the change point month to be April 2020, which coincided with the coronavirus outbreak. The World Health Organization (WHO) announced a global pandemic on March 11, 2020. The New York Times reported on April 2, 2020 that the United States was facing a dire situation, with more than 200,000 confirmed cases in the United States, and the federal medical supplies reserve is almost exhausted.
	
	When $T=200$, the period is from December 2005 to July 2022. The SPACE estimated change point month is September 2008. This month, the financial crisis started to spiral out of control and led to the failure or government takeover of many fairly large financial institutions, triggering an economic recession. 
	
	By comparison, when $T=50, 100, 130$, Dette's estimates are February 2021, January 2021 and October 2020, respectively. However, the implications of the changes are unclear. When $T=150$, Dette's estimator gives the same estimator as SPACE, the time of the coronavirus outbreak. When $T=200$, Detter's estimator also gives the timing of the coronavirus outbreak.

	\section{Conclusion}
	
	In this paper, we propose a break test and a change point location estimator in the correlation structure of high-dimensional data. The main appeal of our methods is that they are developed without any assumption on the ratio of $p$ and $T$ and are thus appropriate for a wide range of large dimensional data sets. The larger $p$ or $T$, the better performance of the proposed methods. Moreover, they are also effective when the change point exists in the extreme tail of a data seqence. Monte Carlo experiments and real data analyses demonstrated the superiority of the proposed methods over several existing methods. In addition, when a change point exists, our methods  automatically provide support recovery in the index set $\boldsymbol{w}_{\tau_1}$ or $\boldsymbol{w}_{\tau_2}$. The proposed estimator is robust to non-Gaussian data, although the theories are developed under the normal assumption. 
	
	There are several possible extensions of our methods. A direct extension is to detect and estimate the change point in the covariance structure of a sequence of high-dimensional vectors. In addition, as we have presented, signflip permutation can break the change point structure and construct the behavior under the null without a change point. Thus, signflip can be used in other change point problems, including online detection. 
	



\appendix
\section{Appendix} 
\subsection{More simulation results}
	$$
	\small
	\begin{array}{ll}
		\hline
		\multicolumn{2}{l}{{\bf Algorithm \ 3}: \text{SPAD+SMOTE}}\\
		\hline
		\ & {\bf Input}: \text{Data Matrix} \ Y_{(i)} \in \mathbb{R}^{p\times T}  (p \ \text{variables and} \ T \ \text{series}), i=1,\ldots,m.\ \text{small positive number} \ \epsilon.\\
		\ & {\bf Output}: \text{the successful detection rate under} \ H_1.\\
		{\bf  \scriptstyle 1} &{\bf for} \ i \leftarrow \text{1 to m \textbf{do}}\\
		{\bf \scriptstyle 2}& \hskip0.6cm a(i) = card(\boldsymbol{w}_{\tau_1})>0,\ a(i)=1\ \text{when there exists a change point},\ a(i)=0,\ \text{otherwise}.\\
		{\bf  \scriptstyle 3} &  {\bf end}\\
		
		{\bf  \scriptstyle 4} & \alpha_0 \leftarrow \frac{sum(a)}{m}, \ \text{the successful detection rate under}\ H_1 \ \text{for raw data.} \\
		
		{\bf  \scriptstyle 5} &{\bf for} \ i \leftarrow \text{1 to m \textbf{do}}\\
		{\bf \scriptstyle 6}& \hskip0.6cm Y_{0(i)} = Y_{(i)}.\\
		{\bf \scriptstyle 7}& \hskip0.6cm  \text{let} \ [\gamma T+1,T] \ \text{columns of} \ Y_{0(i)} \ \text{be the minority class},\ \text{get} (1-\gamma)T \ \text{SMOTE variables},\ \gamma\ge 0.9.\\
		{\bf \scriptstyle 8}& \hskip0.6cm Y_{(i)} \leftarrow  \text{combine} \ Y_{0(i)} \ \text{and SMOTE variables}.\\
		{\bf \scriptstyle 9}& \hskip0.6cm a(i) = card(\boldsymbol{w}_{\tau_1})>0 \ for\  Y_{(i)}. \\
		{\bf  \scriptstyle 10} &  {\bf end}\\
		
		{\bf  \scriptstyle 11} & \alpha_1 \leftarrow \frac{sum(a)}{m}, \ \text{the successful detection rate for inflated data.} \\
		
		{\bf \scriptstyle 12} &{\bf while}\ \alpha_1 - \alpha_0 \ge \epsilon \ {\bf do}\\
		
		{\bf \scriptstyle 13}& \hskip0.6cm \alpha_0 = \alpha_1;\\
		
		{\bf \scriptstyle 14}& \hskip0.6cm \text{Repeat the above steps 5-11.} \\
		{\bf \scriptstyle 15} & {\bf end}\\
		{\bf \scriptstyle 16} & \alpha_1 \leftarrow \text{the successful detection rate under}\ H_1.\\
		\hline
	\end{array}
	$$

	Table~\ref{smote_testing} shows the result of the success rate of detecting the existence of the change point over 200 times repetitions under $H_1$ when the change point occurs in the tail($\beta= 0.9, 0.95$), and the correlation matrix before and after the change point is the same as Case 6 setting. We let $\epsilon=0.05$ in order to improve computing efficiency. Compared with the result of Case 3 in Table~\ref{spad_h1}, where $\beta= 0.75$, when the position of the change point is closer to the tail, the proportion of successful detection of the change point is greatly increased, which also verifies the effectiveness of the iterative algorithm proposed in this paper.

	\begin{table}[h]
		\setlength\tabcolsep{4pt}
		\centering
		\caption{The success rate of detection change point under $H_1$ and iteration time (ITER) of SPAD+SMOTE estimator under Case 6 when the change point exists in the extreme tail, $\epsilon=0.05$, T=100.}
		\footnotesize
		\begin{tabular}{cccccccc}
			\hline
			& $p$ & 20 & 50 & 100 & 200 & 300 & 500 \\ \hline
			\multirow{2}{*}{$\beta=0.9$} & Success rate & 0.8108 & 0.8672 & 0.8961 & 0.9138 & 0.9228 & 0.9327 \\
			& ITER & 3.5650 & 3.2700 & 3.1550 & 3.1150 & 3.0800 & 3.0000 \\ \hline
			\multirow{2}{*}{$\beta=0.95$} & Success rate & 0.6656 & 0.7950 & 0.8492 & 0.8879 & 0.8984 & 0.9082 \\
			& ITER & 3.9800 & 4.0800 & 3.9600 & 3.7900 & 3.7650 & 3.6300 \\ \hline
		\end{tabular}
		\label{smote_testing}
	\end{table}
	
	The following four tables \ref{normal_middle_heter}, \ref{student_middle_heter}, \ref{normal_tail_heter} and \ref{student_tail_heter} contain the simulation results under heteroscedasticity setting in Section 5.4. 
	\begin{table}[h]
		\centering
		\caption{Heteroscedasticity setting. Normal distribution. Mean, standard deviation (SD), and mean squared error (MSE) of three estimators when the change point exists in the middle $\beta=0.5$.}
		\setlength\tabcolsep{4pt}
		\footnotesize
		\begin{threeparttable}[b]
			\begin{tabular}{ccccccccccccc}
				\hline
				\multirow{10}{*}{SPACE} &  & \multicolumn{5}{c}{Case 6} &  & \multicolumn{5}{c}{Case 7} \\ \cline{3-7} \cline{9-13} 
				& $p$ & 5 & 50 & 100 & 300 & 500 & $p$ & 5 & 50 & 100 & 300 & 500 \\
				& Mean & 0.5243 & 0.5231 & 0.5231 & 0.5248 & 0.5217 & Mean & 0.5163 & 0.5227 & 0.5236 & 0.5244 & 0.5201 \\
				& SD & 0.0671 & 0.0290 & 0.0327 & 0.0344 & 0.0312 & SD & 0.1136 & 0.0294 & 0.0326 & 0.0349 & 0.0288 \\
				& MSE & 0.0051 & 0.0014 & 0.0016 & 0.0018 & 0.0014 & MSE & 0.0131 & 0.0014 & 0.0016 & 0.0018 & 0.0012 \\ \cline{2-13} 
				&  & \multicolumn{5}{c}{Case 8} &  & \multicolumn{5}{c}{Case 9} \\ \cline{3-7} \cline{9-13} 
				& $p$ & 5 & 50 & 100 & 300 & 500 & $p$ & 5 & 50 & 100 & 300 & 500 \\
				& Mean & 0.5173 & 0.5165 & 0.5130 & 0.5187 & 0.5160 & Mean & 0.4869 & 0.4761 & 0.4759 & 0.4758 & 0.4760 \\
				& SD & 0.0469 & 0.0241 & 0.0223 & 0.0309 & 0.0249 & SD & 0.0316 & 0.0351 & 0.0365 & 0.0294 & 0.0337 \\
				& MSE & 0.0025 & 0.0008 & 0.0007 & 0.0013 & 0.0009 & MSE & 0.0012 & 0.0018 & 0.0019 & 0.0014 & 0.0017 \\ \hline
				\multirow{10}{*}{\begin{tabular}[c]{@{}c@{}}Dette\\ (2020)\end{tabular}} &  & \multicolumn{5}{c}{Case 6} &  & \multicolumn{5}{c}{Case 7} \\ \cline{3-7} \cline{9-13} 
				& $p$ & 5 & 50 & 100 & 300 & 500 & $p$ & 5 & 50 & 100 & 300 & 500 \\
				& Mean & 0.4947 & 0.5000 & 0.5037 & 0.5067 & 0.4947 & Mean & 0.4130 & 0.4883 & 0.4611 & 0.4840 & 0.4657 \\
				& SD & 0.1209 & 0.0890 & 0.0738 & 0.0682 & 0.0880 & SD & 0.2365 & 0.1202 & 0.1408 & 0.1243 & 0.1417 \\
				& MSE & 0.0146 & 0.0079 & 0.0054 & 0.0047 & 0.0077 & MSE & 0.0632 & 0.0145 & 0.0212 & 0.0156 & 0.0212 \\ \cline{2-13} 
				&  & \multicolumn{5}{c}{Case 8} &  & \multicolumn{5}{c}{Case 9} \\ \cline{3-7} \cline{9-13} 
				& $p$ & 5 & 50 & 100 & 300 & 500 & $p$ & 5 & 50 & 100 & 300 & 500 \\
				& Mean & 0.5076 & 0.5086 & 0.5064 & 0.5165 & 0.5099 & Mean & 0.4845 & 0.4841 & 0.4877 & 0.4882 & 0.4890 \\
				& SD & 0.0797 & 0.0731 & 0.0589 & 0.0379 & 0.0614 & SD & 0.0410 & 0.0429 & 0.0225 & 0.0252 & 0.0416 \\
				& MSE & 0.0064 & 0.0054 & 0.0035 & 0.0017 & 0.0038 & MSE & 0.0019 & 0.0021 & 0.0007 & 0.0008 & 0.0018 \\ \hline
				\multirow{10}{*}{KCP-raw} &  & \multicolumn{5}{c}{Case 6} &  & \multicolumn{5}{c}{Case 7} \\ \cline{3-7} \cline{9-13} 
				& $p$ & 5 & 50 & 100 & 300 & 500 & $p$ & 5 & 50 & 100 & 300 & 500 \\
				& Mean & 0.5238 & 0.6377 & 0.6146 & 0.6262 & 0.6168 & Mean & 0.5645 & 0.6006 & 0.6274 & 0.6689 & 0.6624 \\
				& SD & 0.2903 & 0.1996 & 0.1576 & 0.1547 & 0.1544 & SD & 0.3325 & 0.2812 & 0.2547 & 0.2198 & 0.2054 \\
				& MSE & 0.0844 & 0.0586 & 0.0378 & 0.0397 & 0.0374 & MSE & 0.1141 & 0.0888 & 0.0808 & 0.0766 & 0.0683 \\ \cline{2-13} 
				&  & \multicolumn{5}{c}{Case 8} &  & \multicolumn{5}{c}{Case 9} \\ \cline{3-7} \cline{9-13} 
				& $p$ & 5 & 50 & 100 & 300 & 500 & $p$ & 5 & 50 & 100 & 300 & 500 \\
				& Mean & 0.5381 & 0.6304 & 0.6332 & 0.6837 & 0.6636 & Mean & 0.4726 & 0.3944 & 0.3758 & 0.3655 & 0.3605 \\
				& SD & 0.2965 & 0.2269 & 0.1978 & 0.1771 & 0.1709 & SD & 0.2160 & 0.1788 & 0.1649 & 0.1574 & 0.1695 \\
				& MSE & 0.0889 & 0.0682 & 0.0567 & 0.0650 & 0.0558 & MSE & 0.0472 & 0.0430 & 0.0425 & 0.0427 & 0.0481 \\ \hline
			\end{tabular}
		\end{threeparttable}
		\label{normal_middle_heter}
	\end{table}

	\begin{table}[h]
		\centering
		\caption{Heteroscedasticity setting. Student-t distribution. Mean, standard deviation (SD), and mean squared error (MSE) of three estimators when the change point exists in the middle $\beta=0.5$.}
		\setlength\tabcolsep{4pt}
		\footnotesize
		\begin{threeparttable}[b]
			\begin{tabular}{ccccccccccccc}
				\hline
				\multirow{10}{*}{SPACE} &  & \multicolumn{5}{c}{Case 6} &  & \multicolumn{5}{c}{Case 7} \\ \cline{3-7} \cline{9-13} 
				& $p$ & 5 & 50 & 100 & 300 & 500 & $p$ & 5 & 50 & 100 & 300 & 500 \\
				& Mean & 0.5184 & 0.5460 & 0.5256 & 0.5312 & 0.5279 & Mean & 0.5028 & 0.5265 & 0.5221 & 0.5174 & 0.5223 \\
				& SD & 0.1311 & 0.1050 & 0.1100 & 0.1200 & 0.1129 & SD & 0.2041 & 0.1502 & 0.1413 & 0.1644 & 0.1472 \\
				& MSE & 0.0174 & 0.0131 & 0.0127 & 0.0153 & 0.0135 & MSE & 0.0413 & 0.0232 & 0.0204 & 0.0272 & 0.0220 \\ \cline{2-13} 
				&  & \multicolumn{5}{c}{Case 8} &  & \multicolumn{5}{c}{Case 9} \\ \cline{3-7} \cline{9-13} 
				& $p$ & 5 & 50 & 100 & 300 & 500 & $p$ & 5 & 50 & 100 & 300 & 500 \\
				& Mean & 0.5363 & 0.5343 & 0.5289 & 0.5287 & 0.5233 & Mean & 0.4936 & 0.4728 & 0.4673 & 0.4650 & 0.4680 \\
				& SD & 0.1253 & 0.1042 & 0.1119 & 0.1235 & 0.1054 & SD & 0.0749 & 0.0948 & 0.1030 & 0.1201 & 0.1279 \\
				& MSE & 0.0169 & 0.0120 & 0.0133 & 0.0160 & 0.0116 & MSE & 0.0056 & 0.0097 & 0.0116 & 0.0156 & 0.0173 \\ \hline
				\multirow{10}{*}{\begin{tabular}[c]{@{}c@{}}Dette\\ (2020)\end{tabular}} &  & \multicolumn{5}{c}{Case 6} &  & \multicolumn{5}{c}{Case 7} \\ \cline{3-7} \cline{9-13} 
				& $p$ & 5 & 50 & 100 & 300 & 500 & $p$ & 5 & 50 & 100 & 300 & 500 \\
				& Mean & 0.4430 & 0.2622 & 0.1887 & 0.0909 & 0.0552 & Mean & 0.3173 & 0.1998 & 0.1325 & 0.0658 & 0.0433 \\
				& SD & 0.2088 & 0.2492 & 0.2398 & 0.1777 & 0.1293 & SD & 0.2686 & 0.2502 & 0.2121 & 0.1637 & 0.1114 \\
				& MSE & 0.0466 & 0.1184 & 0.1542 & 0.1987 & 0.2145 & MSE & 0.1052 & 0.1524 & 0.1799 & 0.2152 & 0.2209 \\ \cline{2-13} 
				&  & \multicolumn{5}{c}{Case 8} &  & \multicolumn{5}{c}{Case 9} \\ \cline{3-7} \cline{9-13} 
				& $p$ & 5 & 50 & 100 & 300 & 500 & $p$ & 5 & 50 & 100 & 300 & 500 \\
				& Mean & 0.4472 & 0.3000 & 0.1906 & 0.0866 & 0.0687 & Mean & 0.4852 & 0.3815 & 0.2892 & 0.1903 & 0.1071 \\
				& SD & 0.1955 & 0.2541 & 0.2412 & 0.1773 & 0.1554 & SD & 0.0825 & 0.2147 & 0.2471 & 0.2391 & 0.1887 \\
				& MSE & 0.0408 & 0.1042 & 0.1536 & 0.2022 & 0.2100 & MSE & 0.0070 & 0.0599 & 0.1052 & 0.1528 & 0.1899 \\ \hline
				\multirow{10}{*}{KCP-raw} &  & \multicolumn{5}{c}{Case 6} &  & \multicolumn{5}{c}{Case 7} \\ \cline{3-7} \cline{9-13} 
				& $p$ & 5 & 50 & 100 & 300 & 500 & $p$ & 5 & 50 & 100 & 300 & 500 \\
				& Mean & 0.5292 & 0.5582 & 0.5710 & 0.5812 & 0.5954 & Mean & 0.5461 & 0.5212 & 0.5111 & 0.5814 & 0.5691 \\
				& SD & 0.3020 & 0.2581 & 0.2512 & 0.2483 & 0.2495 & SD & 0.3411 & 0.3481 & 0.3467 & 0.3465 & 0.3516 \\
				& MSE & 0.0916 & 0.0696 & 0.0679 & 0.0679 & 0.0710 & MSE & 0.1179 & 0.1210 & 0.1197 & 0.1261 & 0.1278 \\ \cline{2-13} 
				&  & \multicolumn{5}{c}{Case 8} &  & \multicolumn{5}{c}{Case 9} \\ \cline{3-7} \cline{9-13} 
				& $p$ & 5 & 50 & 100 & 300 & 500 & $p$ & 5 & 50 & 100 & 300 & 500 \\
				& Mean & 0.5460 & 0.5372 & 0.5614 & 0.5951 & 0.5990 & Mean & 0.4992 & 0.4300 & 0.3961 & 0.4392 & 0.4222 \\
				& SD & 0.3089 & 0.3266 & 0.3221 & 0.3162 & 0.2988 & SD & 0.2183 & 0.2403 & 0.2400 & 0.2579 & 0.2627 \\
				& MSE & 0.0971 & 0.1075 & 0.1070 & 0.1085 & 0.0986 & MSE & 0.0474 & 0.0623 & 0.0681 & 0.0699 & 0.0747 \\ \hline
			\end{tabular}
		\end{threeparttable}
		\label{student_middle_heter}
	\end{table}	
	
	\begin{table}[h]
		\centering
		\caption{Heteroscedasticity setting. Normal distribution. Mean, standard deviation (SD), and mean squared error (MSE) of three estimators when the change point exists in the tail $\beta=0.7$.}
		\setlength\tabcolsep{4pt}
		\footnotesize
		\begin{threeparttable}[b]
			\begin{tabular}{ccccccccccccc}
				\hline
				\multirow{10}{*}{SPACE} &  & \multicolumn{5}{c}{Case 6} &  & \multicolumn{5}{c}{Case 7} \\ \cline{3-7} \cline{9-13} 
				& $p$ & 5 & 50 & 100 & 300 & 500 & $p$ & 5 & 50 & 100 & 300 & 500 \\
				& Mean & 0.6315 & 0.6981 & 0.6959 & 0.7015 & 0.6988 & Mean & 0.5744 & 0.6764 & 0.6768 & 0.6861 & 0.6855 \\
				& SD & 0.1072 & 0.0311 & 0.0328 & 0.0265 & 0.0237 & SD & 0.1518 & 0.0530 & 0.0509 & 0.0446 & 0.0419 \\
				& MSE & 0.0161 & 0.0010 & 0.0011 & 0.0007 & 0.0006 & MSE & 0.0387 & 0.0034 & 0.0031 & 0.0022 & 0.0020 \\ \cline{2-13} 
				&  & \multicolumn{5}{c}{Case 8} &  & \multicolumn{5}{c}{Case 9} \\ \cline{3-7} \cline{9-13} 
				& $p$ & 5 & 50 & 100 & 300 & 500 & $p$ & 5 & 50 & 100 & 300 & 500 \\
				& Mean & 0.6690 & 0.7034 & 0.7039 & 0.7032 & 0.7060 & Mean & 0.6585 & 0.6410 & 0.6293 & 0.6223 & 0.6239 \\
				& SD & 0.0738 & 0.0168 & 0.0184 & 0.0115 & 0.0140 & SD & 0.0622 & 0.0693 & 0.0820 & 0.0922 & 0.0878 \\
				& MSE & 0.0064 & 0.0003 & 0.0004 & 0.0001 & 0.0002 & MSE & 0.0056 & 0.0083 & 0.0117 & 0.0145 & 0.0135 \\ \hline
				\multirow{10}{*}{\begin{tabular}[c]{@{}c@{}}Dette\\ (2020)\end{tabular}} &  & \multicolumn{5}{c}{Case 6} &  & \multicolumn{5}{c}{Case 7} \\ \cline{3-7} \cline{9-13} 
				& $p$ & 5 & 50 & 100 & 300 & 500 & $p$ & 5 & 50 & 100 & 300 & 500 \\
				& Mean & 0.5902 & 0.6116 & 0.5772 & 0.5906 & 0.5821 & Mean & 0.4319 & 0.5308 & 0.5011 & 0.5575 & 0.4975 \\
				& SD & 0.1725 & 0.1313 & 0.1667 & 0.1634 & 0.1433 & SD & 0.2590 & 0.2189 & 0.2292 & 0.2030 & 0.2485 \\
				& MSE & 0.0417 & 0.0250 & 0.0427 & 0.0386 & 0.0343 & MSE & 0.1387 & 0.0763 & 0.0918 & 0.0613 & 0.1024 \\ \cline{2-13} 
				&  & \multicolumn{5}{c}{Case 8} &  & \multicolumn{5}{c}{Case 9} \\ \cline{3-7} \cline{9-13} 
				& $p$ & 5 & 50 & 100 & 300 & 500 & $p$ & 5 & 50 & 100 & 300 & 500 \\
				& Mean & 0.6354 & 0.6394 & 0.6381 & 0.6490 & 0.6380 & Mean & 0.6505 & 0.6375 & 0.6484 & 0.6333 & 0.6187 \\
				& SD & 0.1379 & 0.1290 & 0.1196 & 0.0915 & 0.1241 & SD & 0.0796 & 0.1098 & 0.0779 & 0.1132 & 0.1295 \\
				& MSE & 0.0231 & 0.0202 & 0.0181 & 0.0109 & 0.0192 & MSE & 0.0088 & 0.0159 & 0.0087 & 0.0172 & 0.0233 \\ \hline
				\multirow{10}{*}{KCP-raw} &  & \multicolumn{5}{c}{Case 6} &  & \multicolumn{5}{c}{Case 7} \\ \cline{3-7} \cline{9-13} 
				& $p$ & 5 & 50 & 100 & 300 & 500 & $p$ & 5 & 50 & 100 & 300 & 500 \\
				& Mean & 0.6013 & 0.7564 & 0.7768 & 0.7777 & 0.7847 & Mean & 0.5664 & 0.6371 & 0.6759 & 0.7565 & 0.7678 \\
				& SD & 0.3078 & 0.1731 & 0.1412 & 0.1052 & 0.1062 & SD & 0.3363 & 0.3065 & 0.2988 & 0.2159 & 0.2036 \\
				& MSE & 0.1040 & 0.0330 & 0.0257 & 0.0170 & 0.0184 & MSE & 0.1304 & 0.0974 & 0.0894 & 0.0496 & 0.0458 \\ \cline{2-13} 
				&  & \multicolumn{5}{c}{Case 8} &  & \multicolumn{5}{c}{Case 9} \\ \cline{3-7} \cline{9-13} 
				& $p$ & 5 & 50 & 100 & 300 & 500 & $p$ & 5 & 50 & 100 & 300 & 500 \\
				& Mean & 0.5718 & 0.7012 & 0.7408 & 0.8097 & 0.7920 & Mean & 0.6122 & 0.5339 & 0.5192 & 0.5051 & 0.4964 \\
				& SD & 0.3155 & 0.2520 & 0.2243 & 0.1221 & 0.1399 & SD & 0.2382 & 0.2527 & 0.2344 & 0.2329 & 0.2504 \\
				& MSE & 0.1155 & 0.0632 & 0.0517 & 0.0269 & 0.0279 & MSE & 0.0642 & 0.0911 & 0.0873 & 0.0919 & 0.1039 \\ \hline
			\end{tabular}
		\end{threeparttable}
		\label{normal_tail_heter}
	\end{table}

	\begin{table}[h]
		\centering
		\caption{Heteroscedasticity setting. Student-t distribution. Mean, standard deviation (SD), and mean squared error (MSE) of three estimators when the change point exists in the middle $\beta=0.7$.}
		\setlength\tabcolsep{4pt}
		\footnotesize
		\begin{threeparttable}[b]
			\begin{tabular}{ccccccccccccc}
				\hline
				\multirow{10}{*}{SPACE} &  & \multicolumn{5}{c}{Case 6} &  & \multicolumn{5}{c}{Case 7} \\ \cline{3-7} \cline{9-13} 
				& $p$ & 5 & 50 & 100 & 300 & 500 & $p$ & 5 & 50 & 100 & 300 & 500 \\
				& Mean & 0.5954 & 0.6262 & 0.6235 & 0.6138 & 0.6363 & Mean & 0.5393 & 0.5846 & 0.5826 & 0.5602 & 0.5826 \\
				& SD & 0.1710 & 0.1598 & 0.1558 & 0.1778 & 0.1502 & SD & 0.2279 & 0.1902 & 0.1798 & 0.2120 & 0.2000 \\
				& MSE & 0.0400 & 0.0308 & 0.0300 & 0.0389 & 0.0265 & MSE & 0.0773 & 0.0493 & 0.0460 & 0.0643 & 0.0536 \\ \cline{2-13} 
				&  & \multicolumn{5}{c}{Case 8} &  & \multicolumn{5}{c}{Case 9} \\ \cline{3-7} \cline{9-13} 
				& $p$ & 5 & 50 & 100 & 300 & 500 & $p$ & 5 & 50 & 100 & 300 & 500 \\
				& Mean & 0.6353 & 0.6562 & 0.6296 & 0.6329 & 0.6314 & Mean & 0.6321 & 0.5763 & 0.5759 & 0.5628 & 0.5708 \\
				& SD & 0.1584 & 0.1421 & 0.1600 & 0.1755 & 0.1627 & SD & 0.1222 & 0.1593 & 0.1628 & 0.1687 & 0.1649 \\
				& MSE & 0.0292 & 0.0220 & 0.0304 & 0.0352 & 0.0311 & MSE & 0.0195 & 0.0406 & 0.0418 & 0.0472 & 0.0437 \\ \hline
				\multirow{10}{*}{\begin{tabular}[c]{@{}c@{}}Dette\\ (2020)\end{tabular}} &  & \multicolumn{5}{c}{Case 6} &  & \multicolumn{5}{c}{Case 7} \\ \cline{3-7} \cline{9-13} 
				& $p$ & 5 & 50 & 100 & 300 & 500 & $p$ & 5 & 50 & 100 & 300 & 500 \\
				& Mean & 0.4872 & 0.2359 & 0.1740 & 0.0703 & 0.0369 & Mean & 0.3236 & 0.1735 & 0.1201 & 0.0621 & 0.0417 \\
				& SD & 0.2634 & 0.2727 & 0.2600 & 0.1703 & 0.0953 & SD & 0.2819 & 0.2610 & 0.2186 & 0.1671 & 0.1207 \\
				& MSE & 0.1143 & 0.2894 & 0.3440 & 0.4254 & 0.4487 & MSE & 0.2208 & 0.3450 & 0.3839 & 0.4346 & 0.4478 \\ \cline{2-13} 
				&  & \multicolumn{5}{c}{Case 8} &  & \multicolumn{5}{c}{Case 9} \\ \cline{3-7} \cline{9-13} 
				& $p$ & 5 & 50 & 100 & 300 & 500 & $p$ & 5 & 50 & 100 & 300 & 500 \\
				& Mean & 0.5101 & 0.2894 & 0.1586 & 0.0781 & 0.0489 & Mean & 0.6033 & 0.4105 & 0.2842 & 0.1980 & 0.1138 \\
				& SD & 0.2680 & 0.3041 & 0.2568 & 0.1788 & 0.1335 & SD & 0.1604 & 0.3065 & 0.3098 & 0.2897 & 0.2256 \\
				& MSE & 0.1076 & 0.2607 & 0.3588 & 0.4185 & 0.4417 & MSE & 0.0349 & 0.1773 & 0.2684 & 0.3356 & 0.3943 \\ \hline
				\multirow{10}{*}{KCP-raw} &  & \multicolumn{5}{c}{Case 6} &  & \multicolumn{5}{c}{Case 7} \\ \cline{3-7} \cline{9-13} 
				& $p$ & 5 & 50 & 100 & 300 & 500 & $p$ & 5 & 50 & 100 & 300 & 500 \\
				& Mean & 0.5588 & 0.6408 & 0.6515 & 0.6967 & 0.6984 & Mean & 0.5359 & 0.5353 & 0.5593 & 0.6358 & 0.5956 \\
				& SD & 0.3283 & 0.2990 & 0.3048 & 0.2714 & 0.2647 & SD & 0.3454 & 0.3669 & 0.3730 & 0.3625 & 0.3699 \\
				& MSE & 0.1272 & 0.0925 & 0.0948 & 0.0733 & 0.0697 & MSE & 0.1456 & 0.1611 & 0.1583 & 0.1348 & 0.1470 \\ \cline{2-13} 
				&  & \multicolumn{5}{c}{Case 8} &  & \multicolumn{5}{c}{Case 9} \\ \cline{3-7} \cline{9-13} 
				& $p$ & 5 & 50 & 100 & 300 & 500 & $p$ & 5 & 50 & 100 & 300 & 500 \\
				& Mean & 0.5703 & 0.6012 & 0.6035 & 0.6464 & 0.6405 & Mean & 0.6158 & 0.5565 & 0.5083 & 0.5529 & 0.5074 \\
				& SD & 0.3236 & 0.3516 & 0.3554 & 0.3374 & 0.3463 & SD & 0.2619 & 0.2759 & 0.2896 & 0.2977 & 0.3035 \\
				& MSE & 0.1210 & 0.1327 & 0.1350 & 0.1161 & 0.1229 & MSE & 0.0754 & 0.0963 & 0.1202 & 0.1098 & 0.1287 \\ \hline
			\end{tabular}
		\end{threeparttable}
		\label{student_tail_heter}
	\end{table}

	\subsection{Proofs of main results} 
	We first introduce some auxiliary results that are frequently used in the proofs. 
	\begin{lemma1}\label{sub_exp}
		(Lemma 2.7.7 in \cite{vershynin2018}) Consider the samples $\boldsymbol{x}_t=(x_{t1},\ldots,x_{tp})', t=1,\ldots,T$ satisfying Assumption~\ref{A1}. For any $1\le i \le j \le p$, $x_{ti}$ and $x_{tj}$ are sub-Gaussian random variables, then $x_{ti}x_{tj}$ is a sub-exponential random variable.      
	\end{lemma1}
	
	\begin{lemma1}\label{subexp_res1}
		(Proposition 2.7.1 in \cite{vershynin2018}) Let $x_{ti}, t=1,\ldots,T$ be a sub-exponential random variable, then there exist positive constants $c_1,c_2>0$ such that for every $a>0$,
		$$\mathbb{P}(|x_{ti}|\ge a)\le c_1e^{-c_2d}.$$ 
	\end{lemma1}
	\begin{lemma1}\label{subexp_res2}
		(Corollary 2.8.3 in \cite{vershynin2018})   Let $x_{1},\ldots,x_{T} $ be independent, mean zero, sub-exponential random variables, then there exist positive constants $c_1,c_2>0$ such that for every $a>0$,
		$$\mathbb{P}\left\{\left|\frac{1}{T}\sum\limits_{t=1}^Tx_{t}\right|\ge a\right\}\le c_1e^{-c_2T\cdotp \min(a^2,a)}.$$       
	\end{lemma1}     
	
	\subsubsection{Proof of Theorem~\ref{Theorem_1}}
	In Theorem~\ref{Theorem_1}, we prove $\mathbb{P}\left\{\underset{(i,j)\in \mathcal{N}}{\bigcup}\{\boldsymbol{w}(i,j)>\tau_2\}\right\} \rightarrow 0$ under Assumptions~\ref{A1} \& \ref{A3}. Without loss of generality, we can assume $\rho_1(i,j)=\rho_2(i,j)=0$ for any $1\le i, j \le p.$  Observing 
	\begin{eqnarray}\label{T3.1_1}
		\mathbb{P}\left\{\underset{(i,j)\in \mathcal{N}}{\bigcup}\{\boldsymbol{w}(i,j)>\tau_2\}\right\}  \le \underset{(i,j)\in \mathcal{N}}{\sum}\mathbb{P}\{\boldsymbol{w}(i,j)>\tau_2\} \le \frac{1}{2}(p^2-p)\mathbb{P}\{\boldsymbol{w}(i,j)>\tau_2\},  
	\end{eqnarray}
	then it is sufficient to verify $\mathbb{P}\{\boldsymbol{w}(i,j)>\tau_2\}=o\left(\frac{1}{p^2}\right).$

	The components of the vector $\boldsymbol{w}$ corresponding to the entry in the position $(i,j)$ $(1\le i,j\le p)$ of the correlation matrices $\hat{R}_1,
	\hat{R}_2$ are given by 
	$$
	\boldsymbol{w}(i,j)=\frac{1}{T-3} \sum_{t=2}^{T-2} \frac{t (T-t)}{T}  \left(\frac{1}{t} \sum_{k=1}^{t} x_{ki} x_{kj}-\frac{1}{T-t} \sum_{k=t+1}^{T} x_{ki} x_{kj}\right)^{2}
	$$
	
	As $t$ varies from 2 to $T-2$, the weight $\frac{t(T-t)}{T}$ has relatively larger values when $\lfloor \sqrt{T}\rfloor+1 \le t \le T-\lfloor \sqrt{T}\rfloor-1$, compared to its values when $2\le t \le \lfloor \sqrt{T}\rfloor$ and $T-\lfloor \sqrt{T}\rfloor \le t\le T-2$. In addition, when $T$ is big enough, $\sqrt{T}$ can be pretty small, and the two tails of $\frac{t(T-t)}{T}$ show symmetric form. Moreover, both the two tails only involve $(\lfloor \sqrt{T}\rfloor-1)$ terms, and thus the coefficient $\frac{1}{T-3}=\frac{\lfloor \sqrt{T}\rfloor-1}{T-3}\cdotp \frac{1}{\lfloor \sqrt{T}\rfloor-1} $ gives an extra factor of order $\frac{1}{\sqrt{T}}$ in the calculations. Thus, we can decompose $\boldsymbol{w}(i,j)$ in terms  of $t$ into three parts,
	$$
	\boldsymbol{w}(i,j)=\boldsymbol{w}^{(1)}(i,j)+\boldsymbol{w}^{(2)}(i,j)+\boldsymbol{w}^{(3)}(i,j),
	$$
	where $\boldsymbol{w}^{(1)}(i,j)$ corresponds to $t \in \left[2,\lfloor \sqrt{T}\rfloor\right]$, $\boldsymbol{w}^{(2)}(i,j)$ corresponds to $t \in \left[\lfloor \sqrt{T}\rfloor+1, T-\lfloor \sqrt{T}\rfloor-1\right]$, $\boldsymbol{w}^{(3)}(i,j)$ corresponds to $t \in \left[T-\lfloor \sqrt{T}\rfloor, T-2\right]$. Therefore, in order to prove $\frac{1}{2}(p^2-p)\mathbb{P}\{\boldsymbol{w}(i,j)>\tau_2\}\to 0$, it is sufficient to prove
	$$ p^2\mathbb{P}\{\boldsymbol{w}^{(h)}(i,j)>c\tau_2\} \rightarrow 0, h=1,2,3. $$ 
	In this inequality and hereafter in the proof, c and $c_i(i=1,2,\ldots)$ indicate some positive constants that may change from line to line.
	
	\textbf{The case $h=1, 3$}: \\
	For the index $h=1$ and $h=3$ the arguments are quite similar, and for brevity, we only consider the case $h$=1.
	For the statistic $\boldsymbol{w}^{(1)}(i,j)$, we find that
	$$
	\begin{aligned}
		&\mathbb{P}\left\{ \boldsymbol{w}^{(1)}(i,j) \ge c\tau_2  \right\}
		\\& = \mathbb{P}\left\{\frac{\lfloor \sqrt{T}\rfloor-1}{T-3}\cdotp \frac{1}{\lfloor \sqrt{T}\rfloor-1} \sum_{t=2}^{\lfloor \sqrt{T}\rfloor} \frac{t(T-t)}{T}\left(\frac{1}{t} \sum_{k=1}^{t} x_{ki} x_{kj}-\frac{1}{T-t} \sum_{k=t+1}^{T} x_{ki} x_{kj}\right)^{2}> c\tau_2 \right\}\\
		&\leq  \sum_{t=2}^{[ \sqrt{T}]} \mathbb{P} \left\{\left|\frac{1}{t} \sum_{i=k}^{t} x_{ki} x_{kj}-\frac{1}{T-t} \sum_{k=t+1}^{T} x_{ki} x_{kj}\right|>\sqrt{\frac{c \tau_2 T \sqrt{T}}{t(T-t)}}\right\} \\
		&\leq  \sum_{t=2}^{\lfloor \sqrt{T}\rfloor}\left\{ \mathbb{P} \left( \left| \frac{1}{t}\sum_{k=1}^{t} x_{ki} x_{kj} \right| > \sqrt{\frac{c \tau_2 T \sqrt{T}}{t(T-t)}} \right)+  \mathbb{P}\left(\left|\frac{1}{T-t} \sum_{k=t+1}^{T} x_{ki} x_{kj}\right| > \sqrt{ \frac{c \tau_2 T  \sqrt{T} } {t(T-t)}}\right)\right\}.
	\end{aligned}
	$$
	Since $x_{ti}$ and $x_{tj}$ are sub-Gaussian random variables and $x_{ti}x_{tj}$ is a sub-exponential random variable based on Lemma~\ref{sub_exp}. Then, according to Lemmas~\ref{subexp_res1} \& \ref{subexp_res2}, we have
	$$
	\begin{aligned}
		\mathbb{P}\left(\left|\frac{1}{t} \sum_{k=1}^{t} x_{ki} x_{kj}\right|>\sqrt{\frac{c \tau_2  T \sqrt{T}}{t(T-t)}}\right) & \leq c_1 e^{-c_2 t \min \left(\sqrt{\frac{c \tau_2 T \sqrt{T}}{t(T-t)}}, \frac{c \tau_2 T \sqrt{T}}{t(T-t)}\right) }\\
		&\leq c_1 e^{-c_2 \min \left(T^{1/4}\sqrt{\tau_2 }, \tau_2 \sqrt{T}\right)} \leq c_{1} e^{-c_{2} T^{1/4}\sqrt{\tau_2 }},
	\end{aligned}
	$$
	and
	$$
	\begin{aligned}
		\mathbb{P}\left(\left|\frac{1}{T-t} \sum_{k=t+1}^{T} x_{ki} x_{kj}\right|>\sqrt{\frac{c \tau_2 T  \sqrt{T}}{t(T-t)}}\right) &\leq c_1 e^{-c_2(T-t) \min \left(\sqrt{\frac{c \tau_2 T  \sqrt{T}}{t(T-t)}}, \frac{c \tau_2 T \sqrt{T}}{t(T-t)}\right) }\\
		&\leq c_1 e^{-c_2 \min \left(\sqrt{\tau_2 T(T-\sqrt{T})  }, \tau_2 T\right) }\leq c_{1} e^{-c_{2} \sqrt{\tau_2 T(T-\sqrt{T})}},
	\end{aligned}
	$$
	for $t=2,\ldots,\sqrt{T}$.
	Therefore,
	$$
	\mathbb{P}\left\{\boldsymbol{w}^{(1)}(i,j) \ge c\tau_2  \right\}
	\leq  \sum_{t=2}^{\lfloor \sqrt{T}\rfloor} \left(c_{1} e^{-c_{2} \sqrt{\tau_2 \sqrt{T}}}+c_{3} e^{-c_{4} \sqrt{\tau_2 T(T-\sqrt{T})}}\right) \leq c_{1} \sqrt{T} e^{-c_{2} T^{1/4}\sqrt{\tau_2 }},
	$$
	as the smallest absolute value between the exponents is $T^{1/4}\sqrt{\tau_2}$.
	Under Assumption~\ref{A3}, we have
	\begin{eqnarray}\label{part1}
		p^{2} \cdot \mathbb{P}\left\{\boldsymbol{w}^{(1)}(i,j) > c\tau_2 \right\} \leq c_{1} p^{2} \sqrt{T} \cdot e^{-c_{2} T^{1/4}\sqrt{\tau_2}} \rightarrow 0.
	\end{eqnarray}

    \textbf{The case $h=2$}: \\
	For the statistic $\boldsymbol{w}^{(2)}(i,j)$, we have
	$$
	\footnotesize
	\begin{aligned}
		&\mathbb{P}\left\{\boldsymbol{w}^{(2)}(i,j)>c \tau_2\right\}\\
		&=\mathbb{P}\left\{\frac{1}{T-3} \sum_{t=\lfloor \sqrt{T}\rfloor+1}^{T-\lfloor \sqrt{T}\rfloor-1} \frac{t(T-t)}{T}\left(\frac{1}{t} \sum_{k=1}^{t} x_{ki} x_{kj}-\frac{1}{T-t} \sum_{k=t+1}^{T} x_{ki} x_{kj}\right)^{2} > c\tau_2\right\} \\
		&\leq  \sum_{t=\lfloor \sqrt{T}\rfloor+1}^{T-\lfloor \sqrt{T}\rfloor-1} \mathbb{P} \left(\left|\frac{1}{t} \sum_{k=1}^{t} x_{ki} x_{kj}-\frac{1}{T-t} \sum_{k=t+1}^{T} x_{ki} x_{kj}\right| > \sqrt{\frac{c \tau_2 T}{t(T-t)}}\right)  \\
		&\leq  \sum_{t=\lfloor \sqrt{T}\rfloor+1}^{T-\lfloor \sqrt{T}\rfloor-1}  
		\left\{
		\mathbb{P} \left(\left|\frac{1}{t} \sum_{k=1}^{t} x_{ki} x_{kj}\right| > \sqrt{ \frac{c \tau_2 T} {t(T-t)} } \right)   
		+  \mathbb{P} \left( \left| \frac{1}{T-t} \sum_{k=t+1}^{T} x_{ki} x_{kj}\right| > \sqrt{\frac{c \tau_2 T}{t(T-t)}}\right)     \right\}. 
	\end{aligned}
	$$
	Then according to Lemma~\ref{subexp_res2}, we have 
	$$
	\mathbb{P} \left(\left|\frac{1}{t} \sum_{k=1}^{t} x_{ki} x_{kj}\right| > \sqrt{ \frac{c \tau_2 T} {t(T-t)} } \right)  \le c_1 e^{-c_2 t  \min\left(\sqrt{ \frac{c \tau_2 T} {t(T-t)} } ,  \frac{c \tau_2 T} {t(T-t)} \right) }
	\le c_1e^{-c_2 \min(T^{1/4}\sqrt{\tau_2}, \tau_2 ),}
	$$
	and
	$$
	\begin{aligned}
		\mathbb{P} \left( \left| \frac{1}{T-t} \sum_{k=t+1}^{T} x_{ki} x_{kj}\right| > \sqrt{\frac{c \tau_2 T\sqrt{T}}{t(T-t)}}\right)  \le c_1e^{-c_2(T-t)\min\left(\sqrt{\frac{c \tau_2 T}{t(T-t)}} , \frac{c \tau_2 T}{t(T-t)} \right) }  \le c_1e^{-c_2\min\left(T^{1/4}\sqrt{\tau_2}, \tau_2\right)},
	\end{aligned}
	$$
	for $t=\lfloor \sqrt{T}\rfloor+1, \ldots, T-\lfloor \sqrt{T}\rfloor-1 $. Therefore, we have
	$$
	\begin{aligned}
		\mathbb{P}\left(\boldsymbol{w}^{(2)}(i,j) > c\tau_2 \right)
		\le  c_1Te^{-c_2\min(T^{1/4}\sqrt{\tau_2}, \tau_2)}.
	\end{aligned}
	$$
	Under Assumption~\ref{A3}, we have
	$$
	\begin{aligned}
		p^2 \cdot \mathbb{P}\left(\boldsymbol{w}^{(2)}(i,j) > c\tau_2 \right)
		\le  c_1 p^2T \cdot e^{-c_2\min(T^{1/4}\sqrt{\tau_2}, \tau_2)} \rightarrow 0.
	\end{aligned}
	$$
	We complete the proof of Theorem 3.1.

	\subsubsection{Proof of Theorem~\ref{Theorem_2}}
	The result of Theorem~\ref{Theorem_2} is equivalent to
	$$\mathbb{P}\left\{\underset{(i,j)\in \mathcal{P}}{\bigcup}(\boldsymbol{w}(i,j) \le \tau_2)\right\} \rightarrow 0  .$$
	As
	$$\mathbb{P}\left\{\underset{(i,j)\in \mathcal{P}}{\bigcup}(\boldsymbol{w}(i,j) \le \tau_2)\right\} 
	\le \sum_{(i,j)\in \mathcal{P}} \mathbb{P}(\boldsymbol{w}(i,j) \le \tau_2)
	\le p^2\cdot \mathbb{P}(\boldsymbol{w}(i,j) \le \tau_2 ) \rightarrow 0,
	$$
	it is sufficient to prove $p^2\cdot \mathbb{P}(\boldsymbol{w}(i,j) \le \tau_2 ) \rightarrow 0$. A straightforward calculation gives 
	$$ \mathbb{E}\left(\frac{1}{t}\sum_{k=1}^tx_{ki}x_{kj} - \frac{1}{T-t}\sum_{k=t+1}^Tx_{ki}x_{kj}\right)=\left\{
	\begin{aligned}
		\frac{T-t_0}{T-t}\left(\rho_1(i,j)-\rho_2(i,j) \right) ,  t\le t_0\\
		\frac{t_0}{t}\left(\rho_1(i,j)-\rho_2(i,j) \right) ,  t>t_0
	\end{aligned}
	\right.
	$$
	
	Let 
	\begin{eqnarray*}
		A_t&=&\frac{1}{t}\sum_{k=1}^tx_{ki}x_{kj} - \frac{1}{T-t}\sum_{k=t+1}^Tx_{ki}x_{kj} - \mathbb{E}\left(\frac{1}{t}\sum_{k=1}^tx_{ki}x_{kj} - \frac{1}{T-t}\sum_{k=t+1}^Tx_{ki}x_{kj}\right),\\
		B_t &=& \mathbb{E}\left(\frac{1}{t}\sum_{k=1}^tx_{ki}x_{kj} - \frac{1}{T-t}\sum_{k=t+1}^Tx_{ki}x_{kj}\right),
	\end{eqnarray*}
	then $\boldsymbol{w}(i,j) = \frac{1}{T-3}\sum\limits_{t=2}^{T-2}\frac{t(T-t)}{T} (A_t + B_t)^2. $
	When $t\le t_0$, $B_t \ge \frac{T-t_0}{T}(\rho_1(i,j) - \rho_2(i,j) )$, when $t > t_0$, $B_t \ge \frac{t_0}{T}(\rho_1(i,j) - \rho_2(i,j) ) $, so $B_t \ge \min(\frac{T-t_0}{T}, \frac{t_0}{T}) (\rho_1(i,j) - \rho_2(i,j) ) $ and $\frac{1}{|B_t|} \ge \frac{1}{|\rho_1(i,j) - \rho_2(i,j)   | }.$ We have
	\begin{eqnarray}\label{Bt2}
		\begin{aligned}
			\frac{1}{T-3}\sum_{t=2}^{T-2} \frac{t(T-t)}{T} B_t^2 &\ge \frac{1}{T-3}\sum_{t=2}^{T-2} \frac{t(T-t)}{T} \min\left(\frac{(T-t_0)^2}{T^2} , \frac{t_0^2}{T^2}\right) \cdot \left(\rho_1(i,j) - \rho_2(i,j)  \right)^2 \\
			&\ge \min\left(\frac{(T-t_0)^2}{T^2} , \frac{t_0^2}{T^2}\right) \cdot \frac{T\lambda^2}{6}\\
			& \ge \min\left(\frac{(T-t_0)^2}{T^2} , \frac{t_0^2}{T^2}\right) \cdot \frac{T}{6}\cdot c\frac{\tau_2}{T}\cdot \max\left( \frac{T^2}{(T-t_0)^2}, \frac{T^2}{t_0^2} \right)=\frac{c\tau_2}{6},
		\end{aligned}
	\end{eqnarray}
	according to Assumption~\ref{A2}, and then, with a sufficiently large constant $c\ge 12$, we obtain $$\tau_2 \le \frac{1}{2(T-3)} \sum\limits_{t=2}^{T-2} \frac{t(T-t)}{T} B_t^2.$$
	Thus, we can have
	\begin{eqnarray}\label{P_w_2}
		\begin{aligned}
			\mathbb{P}(\boldsymbol{w}(i,j)\le \tau_2) &= 
			\mathbb{P}\left(\frac{1}{T-3}\sum_{t=2}^{T-2}\frac{t(T-t)}{T}(A_t+B_t)^2 \le \tau_2\right)
			\\& \le \mathbb{P}\left( \frac{2}{T-3} \sum_{t=2}^{T-2}\frac{t(T-t)}{T} A_t B_t \le \tau_2 - \frac{1}{T-3}\sum_{t=2}^{T-2}\frac{t(T-t)}{T} B_t^2 \right)
			\\& \le \mathbb{P}\left( \left|\frac{2}{T-3} \sum_{t=2}^{T-2}\frac{t(T-t)}{T} A_t B_t \right| \ge \left| - \frac{1}{2(T-3)}\sum_{t=2}^{T-2}\frac{t(T-t)}{T} B_t^2 \right| \right)
			\\& \le \sum_{t=2}^{T-2} \mathbb{P} \left(|A_t|  \ge  \left|\frac{T}{2t(T-t)}\right| \cdot \frac{1}{|B_t|}  \cdot \frac{1}{2(T-3)}\sum_{t=2}^{T-2}\frac{t(T-t)}{T} B_t^2 \right).
		\end{aligned}
	\end{eqnarray}
	From (\ref{Bt2}), it is easy to obtain $\frac{1}{T-3}\sum\limits_{t=2}^{T-2}\frac{t(T-t)}{T} B_t^2 \ge c\cdot T \min\left(\frac{(T-t_0)^2}{T^2} , \frac{t_0^2}{T^2}  \right) |\rho_1(i,j) - \rho_2(i,j)|^2 $, then the right-hand side within the probability function in (\ref{P_w_2}) is
	\begin{eqnarray*}
		\left|\frac{T}{2t(T-t)}\right| \cdot \frac{1}{|B_t|}  \cdot \frac{1}{2(T-3)}\sum\limits_{t=2}^{T-2}\frac{t(T-t)}{T} B_t^2 &\ge& \frac{cT^2}{t(T-t)} \min\left(\frac{(T-t_0)^2}{T^2} , \frac{t_0^2}{T^2}  \right) |\rho_1(i,j) -\rho_2(i,j)| \\
		&\ge& \frac{cT}{t(T-t)} \sqrt{\tau_2T}.
	\end{eqnarray*}  
	Then $\mathbb{P}(\boldsymbol{w}(i,j)\le \tau_2)$ in (\ref{P_w_2}) becomes
	$$
	\begin{aligned}
		&\mathbb{P}(\boldsymbol{w}(i,j)\le \tau_2)  \le  T \mathbb{P} \left(|A_t|  \ge \frac{cT}{t(T-t)} \sqrt{\tau_2T} \right)
		\\& \le T  \mathbb{P}\left\{ \left|\frac{1}{t}\sum_{k=1}^t x_{ki}x_{kj} -  \frac{1}{T-t}\sum_{k=t+1}^T x_{ki}x_{kj} - 
		\mathbb{E} \left(\frac{1}{t}\sum_{k=1}^t x_{ki}x_{kj} -  \frac{1}{T-t}\sum_{k=t+1}^T x_{ki}x_{kj} \right) \right| \ge \frac{cT}{t(T-t)} \sqrt{\tau_2T}  \right\}
		\\& \le T\mathbb{P}\left\{ \left|\frac{1}{t}\sum_{k=1}^t x_{ki}x_{kj} - \mathbb{E} \left(\frac{1}{t}\sum_{k=t+1}^T x_{ki}x_{kj}\right) \right|  \ge \frac{cT}{t(T-t)} \sqrt{\tau_2T}  \right\}
		\\ & \quad + T\mathbb{P}\left\{ \left|\frac{1}{T-t}\sum_{k=t+1}^T x_{ki}x_{kj} - \mathbb{E} \left(\frac{1}{T-t}\sum_{k=t+1}^T x_{ki}x_{kj}\right) \right|  \ge \frac{cT}{t(T-t)} \sqrt{\tau_2T}  \right\}\\
		& \le T \mathbb{P}\left\{ \left|\frac{1}{t}\sum_{k=1}^t x_{ki}x_{kj} \right| \ge \frac{cT\sqrt{\tau_2T}}{t(T-t)}   \right\}+T \mathbb{P}\left\{ \left|\frac{1}{T-t}\sum_{k=t+1}^T x_{ki}x_{kj} \right| \ge \frac{cT\sqrt{\tau_2T}}{t(T-t)}   \right\}.
	\end{aligned}
	$$
	According to Lemma~\ref{subexp_res2}, for $t \in [2,T-2]$, we have
	$$
	\begin{aligned}
		\mathbb{P}\left\{ \left|\frac{1}{t}\sum_{k=1}^t x_{ki}x_{kj} \right| \ge \frac{cT\sqrt{\tau_2T}}{t(T-t)}   \right\} 
		&\le c_1 e^{-c_2 t \min\left( \frac{cT\sqrt{\tau_2T}}{t(T-t)},  \frac{cT^3\tau_2}{t^2(T-t)^2}\right)  }
		\\&\le c_1e^{-c_2\min(\sqrt{\tau_2T} , \tau_2T )} \le c_1e^{-c_2\sqrt{\tau_2T}},
	\end{aligned}
	$$
	and
	$$
	\begin{aligned}
		\mathbb{P}\left\{ \left|\frac{1}{T-t}\sum_{k=t+1}^T x_{ki}x_{kj} \right| \ge \frac{cT\sqrt{\tau_2T}}{t(T-t)}   \right\} 
		&\le c_1 e^{-c_2 (T-t) \min \left( \frac{cT\sqrt{\tau_2T}}{t(T-t)},  \frac{cT^3\tau_2}{t^2(T-t)^2}\right)  }
		\\&\le c_1e^{-c_2\min(\sqrt{\tau_2T} , \tau_2T )} \le c_1e^{-c_2\sqrt{\tau_2T}},
	\end{aligned}
	$$
	finally, we have $$\mathbb{P}(\boldsymbol{w}(i,j) \le \tau_2) \le c_1Te^{-c_2\sqrt{\tau_2T}}.$$
	Then under Assumption~\ref{A3}, we have $p^2 \cdot \mathbb{P}(\boldsymbol{w}(i,j) \le \tau_2) \le c_1  p^2 Te^{-c_2\sqrt{\tau_2T}} \rightarrow 0,$ we complete the proof.

	\subsubsection{Proof of Theorem~\ref{theorem_3}}
	To prove the consistency of the estimator $\hat{t}$ defined in Equation (\ref{estimation}), we rewrite
	$$\mathbb{P}\left\{ \left|\frac{\hat{\beta}}{\beta}-1\right|\ge \epsilon\right\} = \mathbb{P}\left\{\hat{\beta} \ge (1+\epsilon)\beta \right\}+\mathbb{P}\left\{\hat{\beta}\le (1-\epsilon)\beta\right\}.$$
	We only consider the first term because the second term can be handled similarly. If we want to prove $$\mathbb{P}\{\hat{\beta}\ge \beta(1+\epsilon)\} \rightarrow 0,$$ it is sufficient to show that
	\begin{eqnarray}\label{th3_target}
		\mathbb{P}\left\{\underset{\hat{t}\ge (1+\epsilon)t_0}{\bigcup}\left(U_T(t)\ge U_T(t_0)\right) \right\}  \le  \sum_{t\ge (1+\epsilon)t_0}^T\mathbb{P} \left\{ U_T(t)\ge U_T(t_0)\right\}  \le T\cdotp \mathbb{P}\left(U_T(t)\ge U_T(t_0)\right) \rightarrow 0.
	\end{eqnarray}
	
	Define the vectors $\boldsymbol{m}_k=(m_{k1},\ldots,m_{kd})'=\boldsymbol{z}_k-\mathbb{E}(\boldsymbol{z}_k)$, then
	$$
	\begin{aligned}
		U_T(t)&=\frac{1}{T^4}\sum_{i,k=1}^t \sum_{j,l=t+1}^T\left(\boldsymbol{z}_i - \boldsymbol{z}_j\right)' \left(\boldsymbol{z}_k-\boldsymbol{z}_l\right)
		\\& = \frac{1}{T^4}\sum_{i,k=1}^t \sum_{j,l=t+1}^T
		\left[(\boldsymbol{m}_i-\boldsymbol{m}_j)+(\mathbb{E}(\boldsymbol{z}_i)-\mathbb{E}(\boldsymbol{z}_j))\right]' \left[(\boldsymbol{m}_k-\boldsymbol{m}_l)+(\mathbb{E}(\boldsymbol{z}_k)-\mathbb{E}(\boldsymbol{z}_l))\right]
		\\& =\frac{1}{T^4}\sum_{i,k=1}^t \sum_{j,l=t+1}^T\left(\boldsymbol{m}_i-\boldsymbol{m}_j\right)'\left(\boldsymbol{m}_k-\boldsymbol{m}_l\right) +\frac{1}{T^4}\sum_{i,k=1}^t \sum_{j,l=t+1}^T\left(\mathbb{E}(\boldsymbol{z}_i)-\mathbb{E}(\boldsymbol{z}_j)\right)'\left(\boldsymbol{m}_k-\boldsymbol{m}_l\right)
		\\&\quad +\frac{1}{T^4}\sum_{i,k=1}^t \sum_{j,l=t+1}^T\left(\mathbb{E}(\boldsymbol{z}_k)-\mathbb{E}(\boldsymbol{z}_l)\right)'\left(\boldsymbol{m}_i-\boldsymbol{m}_j\right)+\frac{1}{T^4}\sum_{i,k=1}^t \sum_{j,l=t+1}^T\left(\mathbb{E}(\boldsymbol{z}_i)-\mathbb{E}(\boldsymbol{z}_j)\right)'\left(\mathbb{E}(\boldsymbol{z}_k)-\mathbb{E}(\boldsymbol{z}_l)\right)\\
		&=f(t)+g(t)+h(t)+\delta(t).
	\end{aligned}
	$$
	Let $\boldsymbol{\mu}_1,\boldsymbol{\mu}_2$ be the $d$-dimensional vectors containing the elements of the correlation matrices $R_1$ and $R_2$, respectively, which correspond to positions $(i,j)\in \boldsymbol{w}_{\tau_2}$.
	First, we can calculate
	$$
	\begin{aligned}
		\delta&=\delta(t_0)-\delta(t)
		\\&=\frac{1}{T^4}\sum_{i,k=1}^{t_0} \sum_{j,l=t_0+1}^T\left(\mathbb{E}(\boldsymbol{z}_i)-\mathbb{E}(\boldsymbol{z}_j)\right)'\left(\mathbb{E}(\boldsymbol{z}_k)-\mathbb{E}(\boldsymbol{z}_l)\right) - \frac{1}{T^4}\sum_{i,k=1}^t \sum_{j,l=t+1}^T\left(\mathbb{E}(\boldsymbol{z}_i)-\mathbb{E}(\boldsymbol{z}_j)\right)'\left(\mathbb{E}(\boldsymbol{z}_k)-\mathbb{E}(\boldsymbol{z}_l)\right)
		\\& = \frac{1}{T^4} t_0^2 (t-t_0)(2T-t_0-t)||\boldsymbol{\mu}_1-\boldsymbol{\mu}_2||^2.
	\end{aligned}
	$$
	Then,
	\begin{eqnarray}\label{th3_target_1}
		\mathbb{P}\{U_T(t)\ge U_T(t_0)\} 
		&=& \mathbb{P}\{(f(t)-f(t_0))+(g(t)-g(t_0))+(h(t)-h(t_0))\ge \delta\} 
		\\ \nonumber
		& \le& \mathbb{P}\{f(t)-f(t_0)\ge c\delta\} + \mathbb{P}\{g(t)-g(t_0)\ge c\delta\} +\mathbb{P}\{h(t)-h(t_0)\ge c\delta\}.
	\end{eqnarray}
	
	For the first term in (\ref{th3_target_1}), as
	$$
	\begin{aligned}
		f(t)&=\frac{1}{T^4}\sum_{i,k=1}^t \sum_{j,l=t+1}^T\left(\boldsymbol{m}_i-\boldsymbol{m}_j\right)'\left(\boldsymbol{m}_k-\boldsymbol{m}_l\right)
		\\ &= \frac{(T-t)^2}{T^4}\sum_{i,k=1}^t \boldsymbol{m}_i'\boldsymbol{m}_k - 2\cdot \frac{t(T-t)}{T^4}\sum_{i=1}^t \sum_{l=t+1}^T \boldsymbol{m}_i \mathrm{'} \boldsymbol{m}_l + \frac{t^2}{T^4}\sum_{j,l=t+1}^T \boldsymbol{m}_j' \boldsymbol{m}_l
		\\& = f_1(t) - 2f_2(t) + f_3(t),
	\end{aligned}
	$$
	we have $f(t)-f(t_0)=(f_1(t)-f_1(t_0))-2(f_2(t)-f_2(t_0))+(f_3(t)-f_3(t_0)).$
	For $f_1(t)-f_1(t_0)$, 
	$$
	\begin{aligned}
		&\mathbb{P}(|f_1(t)-f_1(t_0)|\ge c\delta) \\&=
		\mathbb{P}\left( \left| \frac{(T-t)^2}{T^4}\sum\limits_{i,k=1}^t \boldsymbol{m}_i'\boldsymbol{m}_k - \frac{(T-t_0)^2}{T^4}\sum\limits_{i,k=1}^{t_0} \boldsymbol{m}_i'\boldsymbol{m}_k \right| \ge c\delta \right)
		\\& = \mathbb{P}\left( \left| \left(\frac{(T-t)^2}{T^4} - \frac{(T-t_0)^2}{T^4}\right) \sum_{i,k=1}^{t_0} \boldsymbol{m}_i' \boldsymbol{m}_k  
		+\frac{(T-t)^2}{T^4}\sum_{i,k=t_0+1}^t \boldsymbol{m}_i' \boldsymbol{m}_k
		+ 2\cdot \frac{(T-t)^2}{T^4}\sum_{i=1}^{t_0} \sum_{k=t_0+1}^t \boldsymbol{m}_i' \boldsymbol{m}_k   \right|  \ge c\delta \right)\\
		&= \mathbb{P}(-\nu_1+\nu_2+2\nu_3)
	\end{aligned}
	$$
	In order to verify $T\cdotp \mathbb{P}(|f_1(t)-f_1(t_0)\ge c\delta|) \rightarrow 0$, it is sufficient to prove $T\cdotp \mathbb{P}(|\nu_1|\ge c\delta) \rightarrow 0$, $T\cdotp \mathbb{P}(|\nu_2|\ge c\delta) \rightarrow 0$ and $T\cdotp \mathbb{P}(|\nu_3|\ge c\delta) \rightarrow 0.$ First,
	$$
	\begin{aligned}
		\mathbb{P}(|\nu_1|\ge c\delta) 
		&=\mathbb{P}\left(\left| \left(\frac{(T-t_0)^2}{T^4} - \frac{(T-t)^2}{T^4}\right) \sum_{i,k=1}^{t_0} \boldsymbol{m}_i' \boldsymbol{m}_k   \right|  \ge   c\frac{1}{T^4} t_0^2 (t-t_0)(2T-t_0-t)||\boldsymbol{\mu}_1-\boldsymbol{\mu}_2||^2 \right)
		\\& \le \mathbb{P}\left( \left| \frac{1}{d}\sum_{s=1}^d \sum_{i,k=1}^{t_0} m_{is}m_{ks} \right| \ge \frac{ct_0^2}{d}||\boldsymbol{\mu}_1-\boldsymbol{\mu}_2||^2   \right)
		\\& \le \sum_{s=1}^d \mathbb{P}\left( \left| \frac{1}{t_0}\sum_{k=1}^{t_0} m_{ks}^2 \right|  \ge \frac{ct_0}{d}||\boldsymbol{\mu}_1-\boldsymbol{\mu}_2||^2   \right)
		\\&  \le d t_0 \mathbb{P}\left( \left| m_{ks}  \right| \ge c\sqrt{\frac{t_0}{d}}||\boldsymbol{\mu}_1-\boldsymbol{\mu}_2||   \right).
	\end{aligned}  
	$$
	Since $\boldsymbol{m}_k=\boldsymbol{z}_k-\mathbb{E}(\boldsymbol{z}_k)$, and $\boldsymbol{z}_k$ is the $d$-dimensional subvector of $vecho(\boldsymbol{x}_k \boldsymbol{x}_k')$, then $m_{ks}=x_{ks}x_{ks'}-\mathbb{E}(x_{ks}x_{ks'}),$ we can obtain
	$$
	\begin{aligned}
		\mathbb{P}\left( \left| m_{ks}  \right| \ge c\sqrt{\frac{t_0}{d}}||\boldsymbol{\mu}_1-\boldsymbol{\mu}_2||   \right)
		&=  \mathbb{P}\left( \left| x_{ks}x_{ks'}  - \mathbb{E}(x_{ks}x_{ks'} ) \right| \ge c\sqrt{\frac{t_0}{d}}||\boldsymbol{\mu}_1-\boldsymbol{\mu}_2||   \right)
		\\
		&  \le c_1 e^{-c_2 \sqrt{\frac{t_0}{d}}||\boldsymbol{\mu}_1-\boldsymbol{\mu}_2||}  \le c_1e^{-c_2\sqrt{\tau_2}},
	\end{aligned}
	$$
	according to Lemma~\ref{subexp_res1}. Hence,
	\begin{eqnarray}\label{nu1}
		\mathbb{P}(|\nu_1|\ge c\delta) \le  d t_0 c_1e^{-c_2\sqrt{\tau_2}} \le c_1 pTe^{-c_2\sqrt{\tau_2}}.
	\end{eqnarray}
	Second,
	$$
	\begin{aligned}
		&\mathbb{P}(|\nu_2|\ge c\delta)
		\\&=\mathbb{P}\left( \left|\frac{(T-t)^2}{T^4}\sum_{i,k=t_0+1}^t \boldsymbol{m}_i' \boldsymbol{m}_k   \right| \ge c\frac{1}{T^4} t_0^2 (t-t_0)(2T-t_0-t)||\boldsymbol{\mu}_1-\boldsymbol{\mu}_2||^2  \right)
		\\&\le \mathbb{P}\left(\left| \frac{1}{d}\sum_{s=1}^d \sum_{i,k=t_0+1}^t m_{is}m_{ks}  \right| \ge \frac{ct_0^2(t-t_0)}{d(T-t)}||\boldsymbol{\mu}_1-\boldsymbol{\mu}_2||^2  \right)
		\\& \le d \mathbb{P} \left(  \left| \frac{1}{t-t_0} \sum_{k=t_0+1}^t m_{ks}^2  \right| \ge  \frac{ct_0^2}{d(T-t)}||\boldsymbol{\mu}_1-\boldsymbol{\mu}_2||^2 \right)
		\\& \le d(t-t_0) \mathbb{P} \left(  \left|  m_{ki}  \right|  \ge \frac{ct_0 ||\boldsymbol{\mu}_1-\boldsymbol{\mu}_2||}{\sqrt{d(T-t)}} \right),
	\end{aligned}
	$$
	where
	$$
	\begin{aligned}
		&\mathbb{P} \left(  \left|  m_{ki}  \right| \ge  \frac{ct_0 ||\boldsymbol{\mu}_1-\boldsymbol{\mu}_2||}{\sqrt{d(T-t)}} \right) 
		\\& \le \mathbb{P} \left(  \left|  x_{ks}x_{ks'}  -  \mathbb{E}(x_{ks}x_{ks'}) \right| \ge  \frac{ct_0 ||\boldsymbol{\mu}_1-\boldsymbol{\mu}_2||}{\sqrt{d(T-t)}} \right)  
		\\& \le c_1e^{-c_2 \frac{t_0 ||\boldsymbol{\mu}_1-\boldsymbol{\mu}_2||}{\sqrt{d(T-t)}} }  \le c_1e^{-c_2 \sqrt{\tau_2T}} .
	\end{aligned}
	$$
	Hence,
	\begin{eqnarray}\label{nu2}
		\mathbb{P}(|\nu_2|\ge c\delta) \le c_1d(t-t_0)e^{-c_2\sqrt{\tau_2T}} \le c_1 p T e^{-c_2\sqrt{\tau_2T}}.
	\end{eqnarray}
	Third,
	$$
	\begin{aligned}
		&\mathbb{P}(|\nu_3|\ge c\delta)=
		\\& \le \mathbb{P}\left\{ \left|\frac{(T-t)^2}{T^4}\sum_{i=1}^{t_0} \sum_{k=t_0+1}^t \boldsymbol{m}_i' \boldsymbol{m}_k   \right| \ge    c\frac{1}{T^4} t_0^2 (t-t_0)(2T-t_0-t)||\boldsymbol{\mu}_1-\boldsymbol{\mu}_2||^2 \right\}
		\\& \le d \mathbb{P}\left\{ \left| \left(\sum_{k=1}^t m_{ks}  \right)^2 
		- \left(\sum_{k=1}^{t_0} m_{ks} \right)^2  - \left(\sum_{k=t_0+1}^t m_{ks} \right)^2   \right| 
		\ge  \frac{ct_0^2(t-t_0)}{d(T-t)}||\boldsymbol{\mu}_1-\boldsymbol{\mu}_2||^2   \right\}  
		\\& \le d \left\{ \mathbb{P}\left( \left| \frac{1}{t}\sum_{k=1}^t m_{ks}   \right| \ge \frac{ct_0\sqrt{(t-t_0)}||\boldsymbol{\mu}_1-\boldsymbol{\mu}_2||}{t\sqrt{d(T-t)}} \right)  
		+ \mathbb{P}\left( \left|\frac{1}{t_0} \sum_{k=1}^{t_0} m_{ks} \right| \ge \frac{c\sqrt{t-t_0} ||\boldsymbol{\mu}_1-\boldsymbol{\mu}_2||}{\sqrt{d(T-t)}} \right)\right\} 
		\\&\quad + d \mathbb{P}\left( \left| \frac{1}{t-t_0}\sum_{k=t_0+1}^t m_{ks}   \right| \ge \frac{ct_0||\boldsymbol{\mu}_1 - \boldsymbol{\mu}_2||}{\sqrt{d(T-t)(t-t_0)}}   \right),
	\end{aligned}
	$$
	where
	$$
	\begin{aligned}
		&\mathbb{P}\left( \left| \frac{1}{t}\sum_{k=1}^t m_{ks}   \right| \ge \frac{ct_0\sqrt{(t-t_0)}||\boldsymbol{\mu}_1-\boldsymbol{\mu}_2||}{t\sqrt{d(T-t)}} \right)  
		\\ & \le c_1e^{-c_2t \min\left( \frac{ct_0\sqrt{(t-t_0)}||\boldsymbol{\mu}_1-\boldsymbol{\mu}_2||}{t\sqrt{d(T-t)}}  , \frac{ct_0^2(t-t_0)||\boldsymbol{\mu}_1-\boldsymbol{\mu}_2||^2}{dt^2(T-t)}  \right)} \le c_1e^{-c_2\sqrt{\tau_2T}},
	\end{aligned}
	$$
	$$
	\begin{aligned}
		&\mathbb{P}\left( \left| \frac{1}{t_0}\sum_{k=1}^{t_0} m_{ki}   \right| \ge \frac{c\sqrt{(t-t_0)}||\boldsymbol{\mu}_1-\boldsymbol{\mu}_2||}{\sqrt{d(T-t)}} \right)  
		\\& \le c_1e^{-c_2t_0min\left( \frac{c\sqrt{(t-t_0)}||\boldsymbol{\mu}_1-\boldsymbol{\mu}_2||}{\sqrt{d(T-t)}}  , \frac{c(t-t_0)||\boldsymbol{\mu}_1-\boldsymbol{\mu}_2||^2}{d(T-t)}  \right)} \le c_1e^{-c_2\sqrt{\tau_2T}},
	\end{aligned}
	$$
	and
	$$
	\begin{aligned}
		&\mathbb{P}\left( \left| \frac{1}{t-t_0}\sum_{k=t_0+1}^{t} m_{ki}   \right| \ge \frac{ct_0||\boldsymbol{\mu}_1-\boldsymbol{\mu}_2||}{\sqrt{d(T-t)(t-t_0)}} \right)  
		\\& \le c_1e^{-c_2(t-t_0)min\left( \frac{ct_0||\boldsymbol{\mu}_1-\boldsymbol{\mu}_2||}{\sqrt{d(T-t)(t-t_0)}} , \frac{ct_0^2||\boldsymbol{\mu}_1-\boldsymbol{\mu}_2||^2}{d(T-t)(t-t_0)} \right)}\le c_1e^{-c_2\sqrt{\tau_2T}}.
	\end{aligned}
	$$
	Hence, we have
	\begin{eqnarray}\label{nu3}
		\mathbb{P}(|\nu_3|\ge c\delta) \le c_1de^{-c_2\sqrt{\tau_2T}} \le c_1pe^{-c_2\sqrt{\tau_2 T}}.
	\end{eqnarray}
	By combaning (\ref{nu1}), (\ref{nu2}) and (\ref{nu3}), we obtain
	\[\mathbb{P}(|f_1(t)-f_1(t_0)|\ge c\delta) \le c_1 pT e^{-c_2\sqrt{\tau_2}} + c_1 p^2T e^{-c_2\sqrt{\tau_2T}} \le c_1p Te^{-c_2\sqrt{\tau_2}}. \]
	For the other two terms $(f_2(t)-f_2(t_0))$ and $(f_3(t)-f_3(t_0))$ in (\ref{th3_target_1}), following similar analysis, we can obtain
	\begin{eqnarray*}
		&&\mathbb{P}(|f_2(t)-f_2(t_0)|\ge c\delta) \le c_1p Te^{-c_2\sqrt{\tau_2}},\\
		&&\mathbb{P}(|f_3(t)-f_3(t_0)|\ge c\delta) \le c_1p Te^{-c_2\sqrt{\tau_2}}.
	\end{eqnarray*}
	Therefore, we obtain
	\begin{eqnarray}\label{fres}
		\mathbb{P}\{f(t)-f(t_0)\ge c\delta\}\le c_1p Te^{-c_2\sqrt{\tau_2}}.
	\end{eqnarray}
	
	For the second term in (\ref{th3_target}), we have
	$$
	\begin{aligned}
		&g(t)-g(t_0) \\
		&= \frac{1}{T^4}\sum_{i,k=1}^t\sum_{j,l=t+1}^T \left(\mathbb{E}(\boldsymbol{z}_i)-\mathbb{E}(\boldsymbol{z}_j)\right)'(\boldsymbol{m}_k-\boldsymbol{m}_l)  -
		\frac{1}{T^4}\sum_{i,k=1}^{t_0}\sum_{j,l=t_0+1}^T \left(\mathbb{E}(\boldsymbol{z}_i)-\mathbb{E}(\boldsymbol{z}_j)\right)'(\boldsymbol{m}_k-\boldsymbol{m}_l)
		\\& = \frac{1}{T^4}\frac{t_0}{t}(\boldsymbol{\mu}_1-\boldsymbol{\mu}_2)' \left( t(T-t)^2\sum_{k=1}^t \boldsymbol{m}_k - t^2(T-t)\sum_{l=t+1}^T \boldsymbol{m}_l\right)\\
		&\quad - \frac{1}{T^4} (\boldsymbol{\mu}_1 - \boldsymbol{\mu}_2)\mathrm{'} \left( t_0(T-t_0)^2 \sum_{k=1}^{t_0} \boldsymbol{m}_k - t_0^2 (T-t_0) \sum_{l=t_0+1}^T \boldsymbol{m}_l  \right)
		\\& =\frac{t_0}{T^4} (\boldsymbol{\mu}_1-\boldsymbol{\mu}_2)\mathrm{'} \left\{ \left[ (2T-t-t_0)(t_0-t) \right]\sum_{k=1}^{t_0}\boldsymbol{m}_k + \left[ t_0(T-t_0) - t(T-t) \right]\sum_{k=t_0+1}^T \boldsymbol{m}_k  +  T(T-t)  \sum_{k=t_0+1}^t \boldsymbol{m}_k       \right\}\\
		&= - \eta_1+\eta_2+
		\eta_3,
	\end{aligned}
	$$
	where 
	$$
	\begin{aligned}
		& \eta_1=\frac{t_0}{T^4} \left[ (2T-t-t_0)(t-t_0) \right](\boldsymbol{\mu}_1-\boldsymbol{\mu}_2)\mathrm{'}   \sum_{k=1}^{t_0}\boldsymbol{m}_k,
		\\&\eta_2=\frac{t_0}{T^4} \left[ t_0(T-t_0) - t(T-t) \right] (\boldsymbol{\mu}_1-\boldsymbol{\mu}_2)\mathrm{'}  \sum_{k=t_0+1}^T \boldsymbol{m}_k, 
		\\&\eta_3=\frac{t_0}{T^3}   (T-t)(\boldsymbol{\mu}_1-\boldsymbol{\mu}_2)\mathrm{'}   \sum_{k=t_0+1}^t \boldsymbol{m}_k . 
	\end{aligned}
	$$
	To prove $T\cdotp \mathbb{P}(|g(t)-g(t_0)|\ge c\delta) \rightarrow 0, $ it is sufficient to prove $T\cdotp \mathbb{P}(|\eta_1|\ge c\delta) \rightarrow 0$, $T\cdotp \mathbb{P}(|\eta_2|\ge c\delta) \rightarrow 0 $ and $T\cdotp \mathbb{P}(|\eta_2|\ge c\delta) \rightarrow 0 $. First, 
	$$
	\begin{aligned}
		\mathbb{P}(|\eta_1|\ge c\delta)&= \mathbb{P}\left( \frac{t_0}{T^4} \left[ (2T-t-t_0)(t-t_0) \right](\boldsymbol{\mu}_1-\boldsymbol{\mu}_2)\mathrm{'}   \sum_{k=1}^{t_0}\boldsymbol{m}_k  \ge  c\frac{1}{T^4} t_0^2 (t-t_0)(2T-t_0-t)||\boldsymbol{\mu}_1-\boldsymbol{\mu}_2||^2  \right)
		\\& =\mathbb{P}\left( \frac{1}{d} \left| \sum_{s=1}^d \sum_{k=1}^{t_0} m_{ks} \right| \ge \frac{ct_0||\boldsymbol{\mu}_1-\boldsymbol{\mu}_2||}{d}  \right)
		\\& \le d  \mathbb{P}\left( \left| \frac{1}{t_0}\sum_{k=1}^{t_0}  m_{ks} \right|  \ge \frac{c||\boldsymbol{\mu}_1 - \boldsymbol{\mu}_2||}{d}  \right) 
		\\& \le  c_1d e^{-c_2t_0 \min\left(\frac{c||\boldsymbol{\mu}_1-\boldsymbol{\mu}_2||}{d}  , \frac{c||\boldsymbol{\mu}_1-\boldsymbol{\mu}_2||^2}{d^2}  \right)}  \le c_1p^2e^{-c_2 \min(\sqrt{\tau_2T},\tau_2)}.
	\end{aligned}
	$$
	Second, 
	$$
	\begin{aligned}
		&\mathbb{P}(|\eta_2|\ge c\delta)
		\\&=\mathbb{P}\left\{ \frac{t_0}{T^4} \left[ t_0(T-t_0) - t(T-t) \right] (\boldsymbol{\mu}_1-\boldsymbol{\mu}_2)\mathrm{'}  \sum_{k=t_0+1}^T \boldsymbol{m}_k  \ge
		c\frac{1}{T^4} t_0^2 (t-t_0)(2T-t_0-t)||\boldsymbol{\mu}_1-\boldsymbol{\mu}_2||^2 \right\}
		\\& \le \mathbb{P}\left( \frac{1}{d} \left| \sum_{s=1}^d\sum_{k=t_0+1}^T m_{ks}  \right|  \ge \frac{ct_0(T-t_0)||\boldsymbol{\mu}_1-\boldsymbol{\mu}_2||}{d(t_0+t-T)}  \right)
		\\&\le  d  \mathbb{P}\left(\left| \frac{1}{T-t_0}\sum_{k=t_0+1}^T m_{ks}  \right|  \ge  \frac{ct_0||\boldsymbol{\mu}_1-\boldsymbol{\mu}_2||}{d(t_0+t-T)}  \right)
		\\& \le  c_1d e^{-c_2(T-t_0) \min\left(\frac{ct_0||\boldsymbol{\mu}_1-\boldsymbol{\mu}_2||}{d(t_0+t-T)} , \frac{ct_0^2||\boldsymbol{\mu}_1-\boldsymbol{\mu}_2||^2}{d^2(t_0+t-T)^2}\right)  }   \le c_1p^2 e^{-c_2 \min(\sqrt{\tau_2T},\tau_2)}.
	\end{aligned}
	$$
	Last,
	$$
	\begin{aligned}
		\mathbb{P}(|\eta_3|\ge c\delta) &=\mathbb{P}\left\{\frac{t_0}{T^3}   (T-t)(\boldsymbol{\mu}_1-\boldsymbol{\mu}_2)\mathrm{'}   \sum_{k=t_0+1}^t \boldsymbol{m}_k      \ge c\frac{1}{T^4} t_0^2 (t-t_0)(2T-t_0-t)||\boldsymbol{\mu}_1-\boldsymbol{\mu}_2||^2 \right\}
		\\& \le \mathbb{P}\left( \left|\frac{1}{d}\sum_{s=1}^d\sum_{k=t_0+1}^t m_{ks}   \right| \ge \frac{ct_0(t-t_0)||\boldsymbol{\mu}_1-\boldsymbol{\mu}_2||}{dT}  \right)
		\\& \le d  \mathbb{P} \left( \left| \frac{1}{t-t_0}\sum_{k=t_0+1}^t m_{ks} \right| \ge \frac{ct_0||\boldsymbol{\mu}_1-\boldsymbol{\mu}_2||}{dT}  \right)   
		\\&\le c_1p^2e^{-c_2 \min(\sqrt{\tau_2T},\tau_2)}.
	\end{aligned}
	$$
	Therefore, we obtain
	\begin{eqnarray}\label{gres}
		\mathbb{P}(|g(t)-g(t_0)|\ge c\delta)\le c_1p^2e^{-c_2 \min(\sqrt{\tau_2T},\tau_2)}.
	\end{eqnarray}
	For the third term in (\ref{th3_target_1}), similarly, we can obtain
	\begin{eqnarray}\label{hres}
		\mathbb{P}(|h(t)-h(t_0)|\ge c\delta)\le c_1p^2 e^{-c_2 \min(\sqrt{\tau_2T},\tau_2)}.
	\end{eqnarray} 
	By combining (\ref{fres}), (\ref{gres}), and (\ref{hres}), we can prove that  
	$\mathbb{P}(U_T(t)\ge U_T(t_0)) 
	\le c_1p^2e^{-c_2\sqrt{\tau_2}}$ and Equation (\ref{th3_target}) holds under Assumption~\ref{A3}. We complete the proof.


\bibliographystyle{imsart-nameyear} 
\bibliography{refs}      


\end{document}